\documentclass[a4paper,fleqn,usenatbib]{mnras}

\usepackage{newtxtext,newtxmath}

\usepackage[T1]{fontenc}
\usepackage{ae,aecompl}

\usepackage{graphicx}	
\usepackage{amsmath}	
\usepackage{amssymb}	

\title[Axisymmetric density waves]{Axisymmetric density waves in Saturn's rings}
\usepackage{placeins}
\usepackage{natbib}
\usepackage{aasref}
\author[M.M. Hedman and P.D. Nicholson ]{
M.M. Hedman,$^{1}$\thanks{E-mail: mhedman@uidaho.edu (MMH)}
and P.D. Nicholson$^{2}$
\\
$^{1}$Department of Physics, University of Idaho, Moscow ID  USA 83844-0903 \\
$^{2}$Department of Astronomy, Cornell University, Ithaca NY USA 14853
}

\date{Accepted XXX. Received YYY; in original form ZZZ}
\pubyear{2019}

\begin{document}
\label{firstpage}
\pagerange{\pageref{firstpage}--\pageref{lastpage}}
\maketitle



\begin{abstract}
Density waves in Saturn's rings are usually tightly wrapped spiral patterns generated by resonances with either Saturn's moons or structures inside the planet. However, between the Barnard and Bessel Gaps in the Cassini Division (i.e. between 120,240 and 120,300 km from Saturn's spin axis), there are density variations that appear to form an axisymmetric density wave consisting of concentric zones of varying densities that propagate radially through the rings. Axisymmetric waves cannot be generated directly by a satellite resonance, but instead appear to be  excited by interference between a nearby satellite resonance and normal mode oscillations on the inner edge of the Barnard Gap. Similar axisymmetric waves may exist just interior to other resonantly confined edges that exhibit a large number of normal modes, including the Dawes ringlet in the outer C ring and the outermost part of the B ring.
\end{abstract}

\maketitle

\section{Introduction}
\label{intro}

Among the best understood class of features in Saturn's rings are density waves. These structures are usually tightly wound multi-armed spiral patterns that are generated at mean-motion resonances with either Saturn's moons or structures in the planet's interior. The theory behind these patterns is very well developed \citep{GT82, Shu84}, enabling ring parameters like the local surface mass density to be derived from observable wave properties \citep{Cuzzi81, Esposito83, Tiscareno07, Esposito10}. At the same time, the predictable properties of these waves allow wavelet-based filtering techniques to identify wave-like structures that are not apparent in individual observations \citep{HN16}. These tools are also starting to reveal unexpected waves with unusual properties, including several features that appear to be {\em axisymmetric} density waves (i.e.\ waves with azimuthal wavenumber $m=0$).

The clearest example of an axisymmetric density wave lies between the Bessel and Barnard Gaps in the Cassini Division. This region had been designated the 1.994 $R_s$ ringlet prior to the Cassini mission \citep{NCP90, French93} when the two gaps were thought to be a single gap inhabited by a ringlet nearly as wide as the gap. However, Cassini observations later revealed that the Bessel and Barnard gaps likely had different dynamical origins \citep{Colwell09, French10, Hedman10, French16} and so it seemed more appropriate to consider them as separate gaps. Hence the region in between them was not given a formal name.  
\pagebreak

\begin{figure}
\centerline{\resizebox{3.5in}{!}{\includegraphics{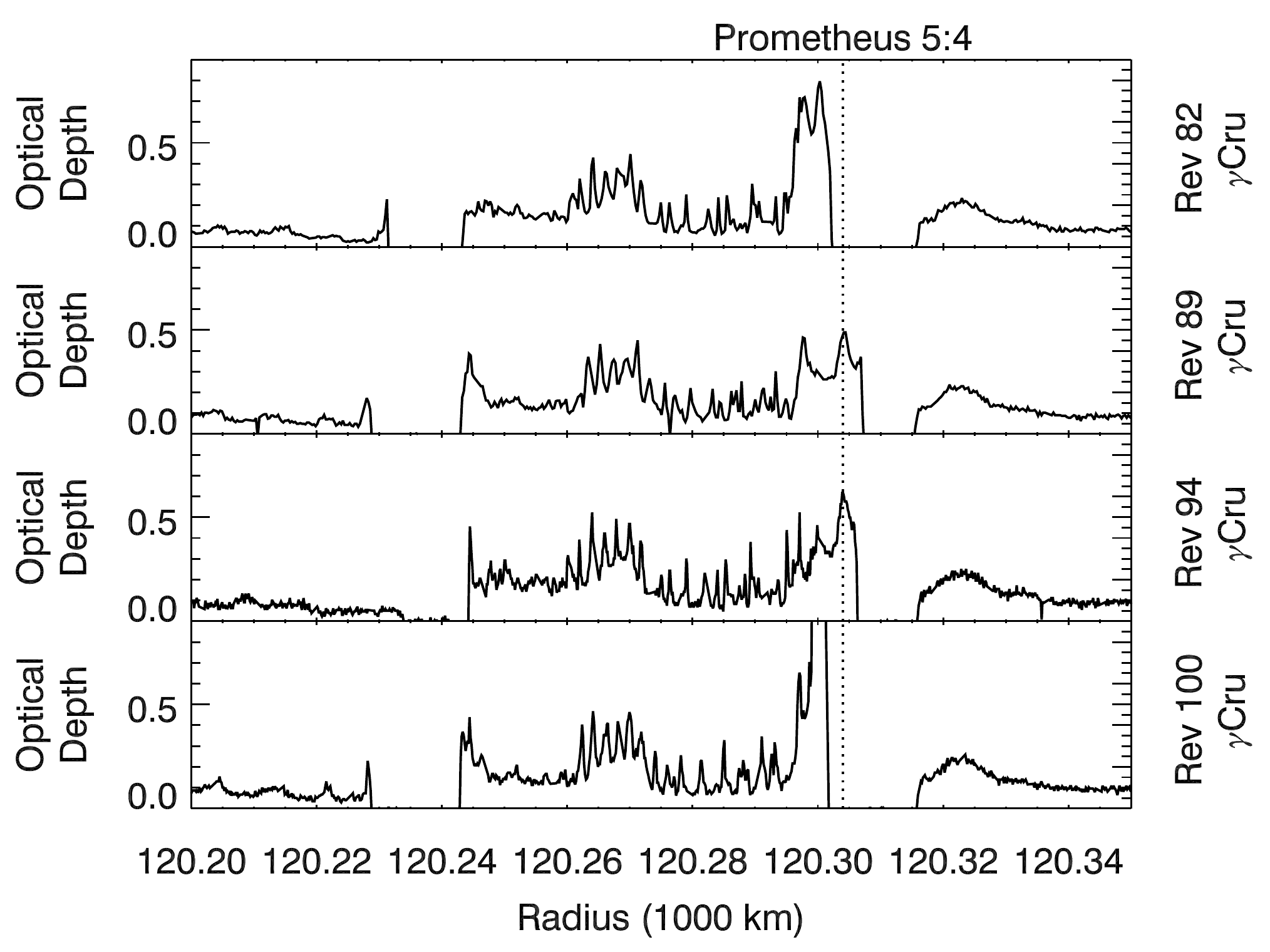}}}
\caption{A few representative optical depth profiles of the region between the Bessel and Barnard Gaps derived from stellar occultations obtained by the VIMS instrument onboard the Cassini spacecraft.}
\label{baedge}
\end{figure}

As shown in Figure~\ref{baedge}, the region between the Bessel and Barnard gaps contains quasi-periodic optical depth variations and the locations of the peaks and troughs vary from one observation to another. This is reminiscent of typical density waves, but this particular structure is anomalous in that there is no resonance with any of Saturn's moons that would naturally generate this wave. While the Prometheus 5:4 inner Lindblad resonance lies at 120,304 km (near the inner edge of the Barnard gap) the wave generated by this resonance should propagate outwards, away from the region of interest. \citep[In fact, this resonance instead excites a five-lobed radial variation in the edge's position, see ][]{French16}.

A wavelet-based analysis of this structure provides strong evidence that it is an axisymmetric density wave consisting of concentric zones of varying density. Such a wave is a valid solution to the relevant equations of motion for ring material, and the standard theory of density waves can naturally be extended to this case, yielding sensible estimates of the local surface mass density. However, such a structure is not easily excited by any mean-motion resonance with a satellite because the gravitational perturbations from a moon must always vary with longitude. An axisymmetric wave instead requires some process that induces all the particles to have finite orbital eccentricities and to pass through pericenter at the same time regardless of longitude. We propose that  interference among the observed normal-mode oscillations in the position of the Barnard gap's inner edge can give rise to perturbations with the correct form to generate an axisymmetric density wave.

\begin{figure}
\centerline{\resizebox{3.5in}{!}{\includegraphics{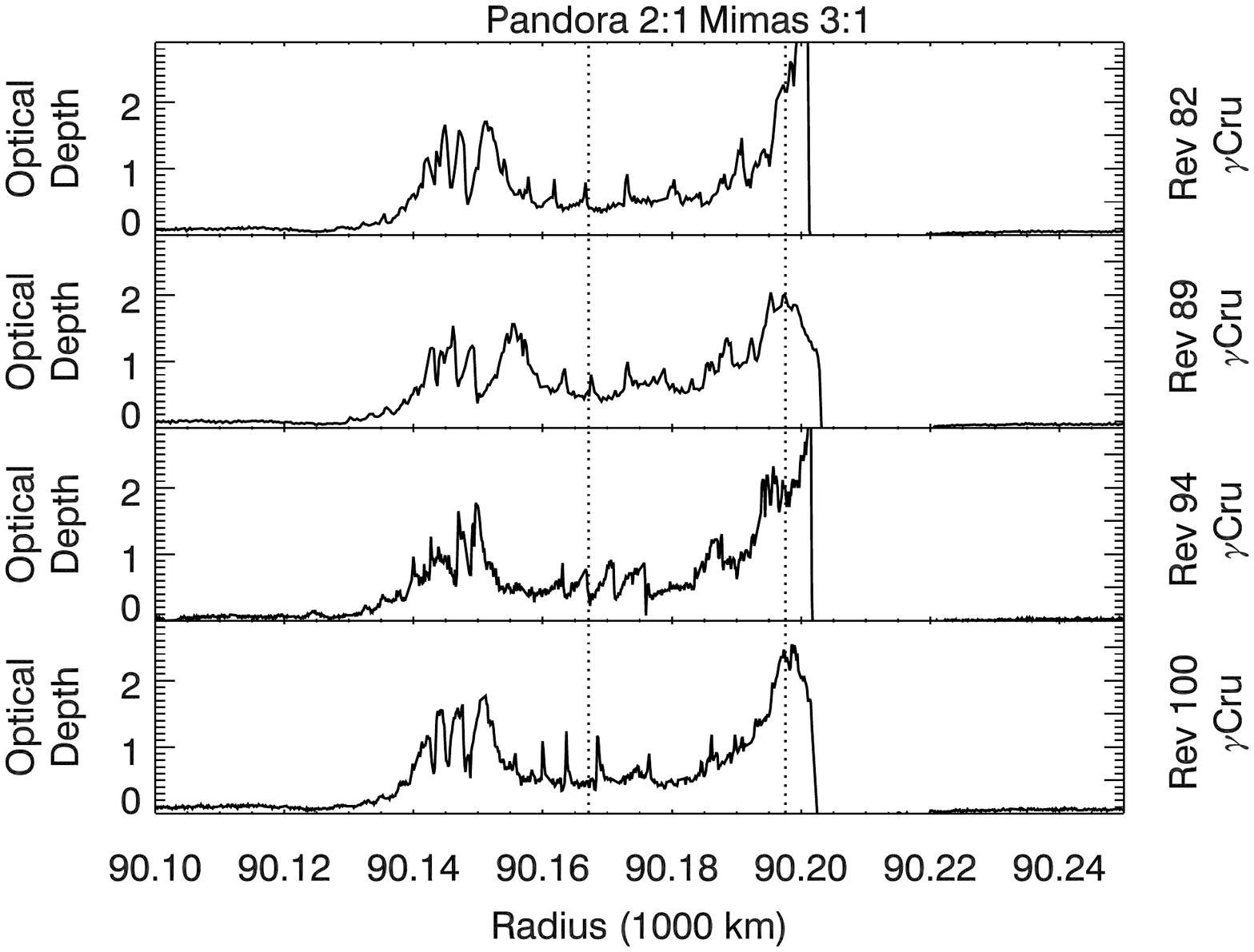}}}
\caption{A few representative optical depth profiles of the  Dawes ringlet derived from stellar occultations obtained by the VIMS instrument onboard the Cassini spacecraft.}
\label{dawes}
\end{figure}

\begin{figure}
\centerline{\resizebox{3.5in}{!}{\includegraphics{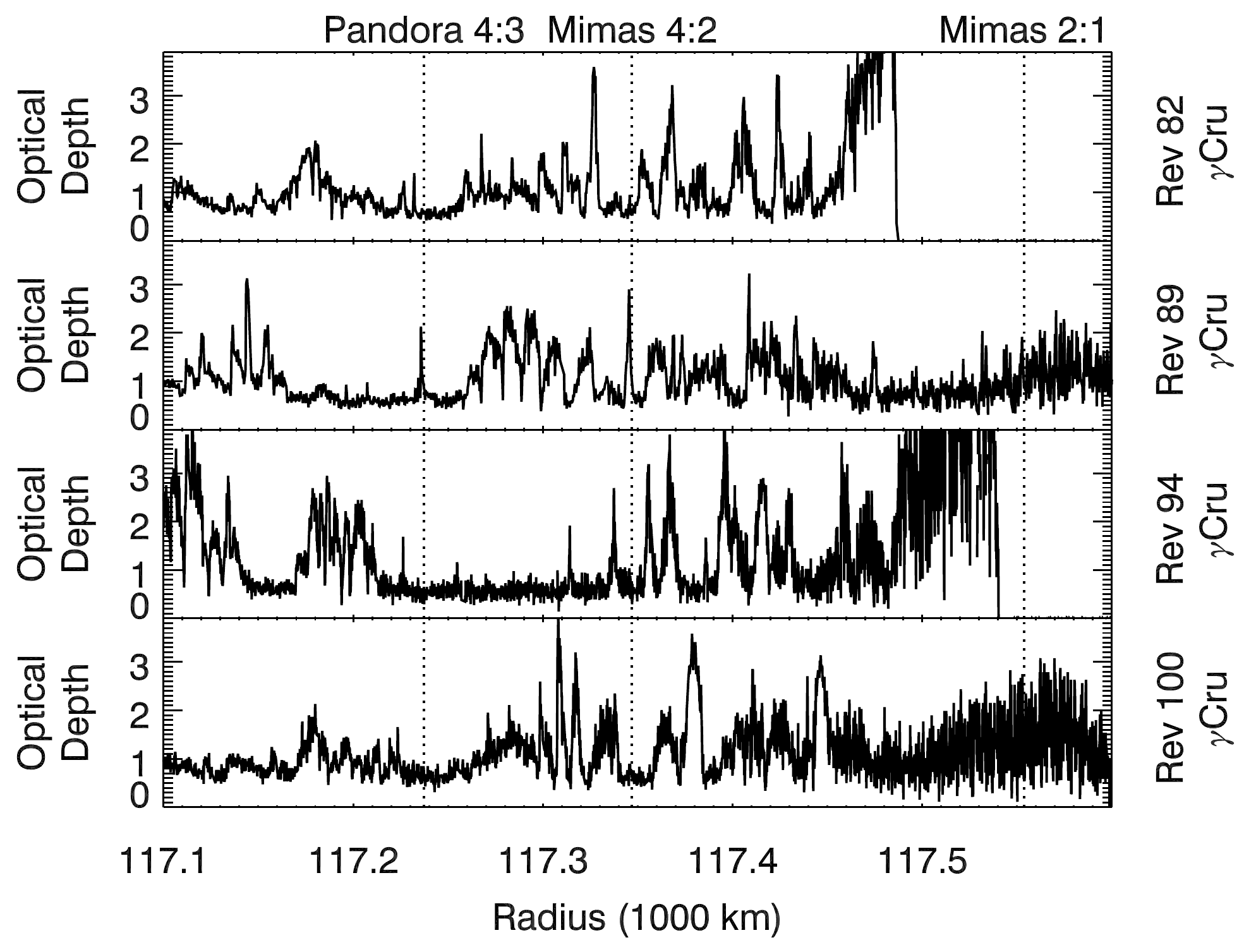}}}
\caption{A few representative optical depth profiles of the B ring outer edge derived from stellar occultations obtained by the VIMS instrument onboard the Cassini spacecraft.}
\label{bedge}
\end{figure}
 
 This wave would not be the first axisymmetric structure to be found in planetary rings. Prior analyses of occultation data for Uranus' rings revealed that the $\gamma$ ring exhibits axisymmetric variations in its radial position \citep{French91}. The $\gamma$ ring is also found close to a mean motion resonance with one of Uranus' moons (specifically, the 6:5 inner Lindblad resonance with Ophelia), so the dynamical environment of the $\gamma$ ring is similar to that of the material around the Barnard Gap. The axisymmetric motion of Uranus' $\gamma$ ring and the axisymmetric wave in the Cassini Division could therefore represent a previously unrecognized class of ring features that are excited in the vicinity of resonantly-confined ring material.

Indeed, other resonantly-confined edges of Saturn's rings, like  the outer edges of the B ring and the Dawes ringlet (which are perturbed by the Mimas 2:1 and 3:1 inner Lindblad resonances, respectively) are associated with periodic optical depth variations that may be additional examples of axisymmetric density waves.  Unfortunately, the wavelet-based techniques that could characterize the patterns near the Barnard gap do not appear to work on these structures, making their interpretation less certain.

Figure~\ref{dawes} shows that the Dawes ringlet contains periodic opacity variations of similar scale to those seen interior to the Barnard Gap. These variations also extend interior to the Mimas 3:1 and Pandora 2:1 resonances in this region, and so cannot just be density waves driven by those satellites (which would propagate outwards). This feature might be an axisymmetric density wave, but other structures in the region complicate its interpretation.

The outer part of the B ring is far more complex (see Figure~\ref{bedge}), and is perturbed by a number of strong resonances. However, some quasi-periodic signals can be seen around 117,300 km. These are not in the right place to be generated by the known satellite resonances. Furthermore, imaging data support the idea that these structures are another example of an axisymmetric density wave (see Section~\ref{bim} below). 

The rest of this paper provides a detailed analysis of these structures. Section~\ref{background} provides the relevant theoretical background for density waves in order to clarify the expected properties of axisymmetric waves and how they can be distinguished from other ring features. Section~\ref{barnard} describes our analysis of the region between the Bessel and Barnard gaps and the evidence that there is an axisymmetric density wave in this region.  Section~\ref{model} then discusses how this sort of axisymmetric wave could be excited by interference among the various normal modes on resonantly-confined gap edges. Section~\ref{others} examines the material in the vicinity of other resonantly confined gap edges and whether these might also contain axisymmetric density waves. Finally, Section~\ref{summary} summarizes the results and implications of these analyses.

\section{Theoretical Background}
\label{background}

The theory behind spiral density waves is described in several classic papers \citep[e.g.][]{Shu84}. However, for the sake of convenience we will review the relevant equations here in order to clarify how axisymmetric waves should behave.

Generically, density waves arise when a dense ring of material is subjected to a periodic perturbing force that generates a term $\mathcal{U}$ in the potential that varies periodically with time $t$ and/or inertial longitude $\lambda$: 

\begin{equation}
\mathcal{U} \propto \exp[i(|m|\lambda-\omega t)],
\label{perturb}
\end{equation} 
where $\omega$ is the perturbation frequency and $m$ is a integer that can be positive, negative or 0, with $m=0$ corresponding to an axisymmetric perturbation\footnote{Note that the sign convention used here for the phase parameter in the exponential is the opposite of that used in \citet{Shu84}}. This sort of perturbation has its strongest effect on the orbital motion of the ring particles at locations where the ring-particles' radial epicyclic frequency $\kappa$ and orbital mean motion $n$ satisfy this relationship (for either choice of sign):

\begin{equation}
\omega=|m|n\mp \kappa,
\label{resrel}
\end{equation}
At these locations, we can re-write the frequency $\omega$ in Equation~\ref{perturb} in terms of the mean motion and radial epicyclic frequency. Furthermore, for each ring particle $\lambda-nt$ is a constant, and so the perturbation has the following form:

\begin{equation}\mathcal{U} \propto
\exp[i(|m|\lambda-|m|nt\pm\kappa t)]=\exp[\pm i\kappa t].\end{equation} The perturbations on each particle therefore have the same frequency as the particle's radial epicyclic motion, which is the appropriate condition for a Lindblad resonance that will excite and organize the ring particles' epicyclic motions. Furthermore, if the ring has a finite surface mass density $\sigma$ where this resonance occurs, then these localized disturbances will propagate radially across the rings, forming the pattern of surface mass density variations known as a density wave. 

In general, the density variations associated with any density wave can be quantified using the following expression.

\begin{equation}
\sigma =\sigma_0 + \Re\left[A_\sigma e^{i(\phi_r+\phi_{\lambda t})}\right],
\end{equation}
where $\sigma_0$ is the background surface mass density and $A_\sigma$ is a (radius-dependent) amplitude of the density variations, while $\phi_{\lambda t}$ and $\phi_r$ are phase parameters that depend on longitude/time and radius, respectively. The functional forms of these two phases are determined by the characteristics of the perturbation and the ring's background surface mass density.

First, consider the longitude/time dependent phase $\phi_{\lambda t}$. Steady state solutions for the density variations only exist if the pattern as a whole tracks the potential, so $\phi_{\lambda t}$ must have the same form as the phase in the perturbing potential shown in Equation~\ref{perturb}:

\begin{equation}
\phi_{\lambda t}=|m|\lambda -\omega t
\label{phidef}
\end{equation}
The entire pattern therefore has azimuthal wavenumber $|m|$. Also, if $m \ne 0$, then this pattern rotates at a speed $\Omega_p=\omega/|m|$, which can be written in terms of the mean motion and radial epicyclic frequency at the resonant location $n_L$ and $\kappa_L$:

\begin{equation}
\Omega_p(m\ne 0)=n_L\mp\frac{1}{|m|}\kappa_L=n_L-\frac{1}{m}\kappa_L
\label{eqpat}
\end{equation}
where for the second equality we have chosen the sign of $m$ such that positive $m$  correspond to situations where the pattern speed is slower than the orbital mean motion (i.e. an Inner Lindblad Resonance) while negative $m$ correspond to cases where the pattern speed is faster than the local orbital mean motion (i.e. an Outer Lindblad Resonance). Of course, for the axisymmetric case ($m=0$), there is no sensible definition of a real azimuthal pattern speed. However, since  in this case $\phi_{\lambda t}=-\omega t$, a reasonable analog of the pattern speed is 

\begin{equation}
\Omega_p(m =0)=\omega=\kappa_L
\label{eqpat0}
\end{equation}
where the second equality uses the general definition of $\omega$, and we have chosen the positive sign in order to ensure that $\Omega_p$ and $\omega$ are sensibly positive. 


While $\phi_{\lambda t}$ depends only on the properties of the perturbation, the radius-dependent part of the wave phase $\phi_r$ is determined by a dispersion relation that relates the wave frequency $\omega$ to  the radial wavenumber of the pattern $k$, which is simply the radial derivative of the wave phase at a fixed longitude and time:

\begin{equation}
k(r)=\frac{\partial \phi_r}{\partial r}.
\end{equation}
In the limit where the velocity dispersion of the ring particles is negligible (which is appropriate for the ring regions considered here), the relevant dispersion relation is \citep{Shu84}:

\begin{equation}
(\omega-|m|n)^2=\kappa^2-2\pi G\sigma_0|k|.
\label{dispersion}
\end{equation}
So long as the wave is observed at radii $r$ close to the radius of the resonance $r_L$ (i.e $|r-r_L|<<r_L$) and the apsidal precession rate is much less than the mean motion (i.e. $\kappa \simeq n$), this expression yields the standard expression for the wavenumber of a wave (valid for all $m \ne 1$):

\begin{equation}
|k(r)|=\frac{3(m-1) M_P(r-r_L)}{2\pi\sigma_0r_L^4},
\label{keq}
\end{equation}
where $M_P$ is the mass of the planet. Note that if $m>1$ then the right hand side of this equation is only positive if  $r>r_L$, which is consistent with the fact that such density waves only exist outside  Inner Lindblad Resonances. By contrast, if $m<1$ (including $m=0$), then the wave only exists interior to the resonance.  Furthermore, in order for the waves to propagate away from the resonant location in the appropriate direction, their group velocity $v_g=\partial \omega /\partial k$ also needs to be positive if $m>1$ and negative if $m<1$. Taking the appropriate derivatives of Equation~\ref{dispersion} reveals that these conditions will be satisfied provided $k$ is positive  (again, opposite the Shu 1984 convention). Note that since $m=0$ waves behave like those generated by standard Outer Lindblad Resonances in this regard, it is sensible to regard $m=0$ resonances as a member of that group.\footnote{Note also that the dispersion relation in Equation~\ref{dispersion} implies that  $m=0$ waves can only propagate where
$\kappa > |\omega|$, i.e., interior to the Lindblad resonance}  Also, for all $m$, the magnitude of the group velocity is given by the standard expression:

\begin{equation}
|v_g|=\frac{\pi G\sigma_0}{\kappa_L}.
\label{group}
\end{equation}

Since $k=\partial \phi_r/\partial r$ is positive for both types of waves, the phase always increases with increasing radius, which is consistent with how phases are defined for the wavelet transformations (see below). Furthermore, at a fixed longitude, the location of maxima or minima will drift outwards over time for all $m$. In other words, the phase velocity $v_p=\omega/k$ is positive definite for both types of waves, consistent with all the waves with $m\ne 0$ being trailing spiral patterns.  Finally, we can integrate Equation~\ref{keq} for $|k|=k$ to obtain the following expression for the radial phase parameter in the vicinity of the resonance:

\begin{equation}
\phi_r(r)=\frac{3(m-1) M_P(r-r_L)^2}{4\pi\sigma_0 r_L^4} +\phi_0
\label{phir}
\end{equation}
where $\phi_0$ is a constant phase offset. 

In summary, standard density wave theory predicts that axisymmetric $m=0$ density waves should have the following properties:
\begin{itemize}
\item They should exist interior to the resonant location where $\omega=\kappa$. 
\item They should have negative group velocities, which means they should propagate and carry energy and angular momentum inwards.
\item They should have positive phase velocities, which means the positions of individual peaks and troughs should move outwards over time.
\item They should have radial wavenumbers  given by the expression $k(r)=\frac{3 M_P|r-r_L|}{2\pi\sigma_0r_L^4}$
\end{itemize}

\section{Analysis of the region between the Barnard and Bessel Gaps}
\label{barnard}

The region between the Barnard and Bessel Gaps shown in Figure~\ref{baedge} contains the feature that can be most convincingly identified as an axisymmetric density wave based on wavelet analysis of multiple occultation profiles. The relevant occultation data are described in Section~\ref{obs}, while Section~\ref{wave} provides the wavelet-based methods employed to characterize this particular pattern. Finally, Section~\ref{results} presents the resulting evidence that the region interior to the Barnard Gap does indeed contain an axisymmetric wave.

\begin{table*}
\caption{Summary of occultations used in this study}
\label{occs}
\centerline{\resizebox{6.5in}{!}{
\begin{tabular}{|c|c|c|c||c|c|c|} \hline
Star & Rev$^a$ & i/e$^b$ & B$^c$ &  ET$^d$ &  Longitude$^e$ & m=0 Phase$^f$  \\ & & & (degrees) & (seconds) & (degrees) & (degrees) \\ \hline
RHya & 036 & i &  -29.4 &  220943101. &  195.4  &  349.6 \\ 
alpSco & 013 & i &  -32.2 &  177810304. &  288.6  &  324.7 \\ 
alpSco & 013 & e &  -32.2 &  177817667. &  358.0  &  262.6 \\ 
alpSco & 029 & i &  -32.2 &  212528274. &  194.6  &  359.9 \\ 
alpAur & 041 & i &   50.9 &  227942859. &    8.9  &   27.2 \\ 
RCas & 065 & i &   56.0 &  262010525. &   29.6  &  144.0 \\ 
gamCru & 071 & i &  -62.3 &  266187755. &  186.7  &  213.3 \\ 
gamCru & 072 & i &  -62.3 &  266804308. &  186.3  &   56.2 \\ 
gamCru & 073 & i &  -62.3 &  267420458. &  186.0  &  262.6 \\ 
gamCru & 077 & i &  -62.3 &  269852643. &  185.1  &  281.2 \\ 
gamCru & 078 & i &  -62.3 &  270461132. &  184.9  &  192.1 \\ 
betGru & 078 & i &  -43.4 &  270509701. &  263.3  &  142.7 \\ 
RSCnc & 080 & i &   30.0 &  271864841. &   56.1  &  240.0 \\ 
RSCnc & 080 & e &   30.0 &  271884854. &  155.6  &   71.3 \\ 
gamCru & 081 & i &  -62.3 &  272314456. &  183.2  &   50.1 \\ 
gamCru & 082 & i &  -62.3 &  272950229. &  182.9  &   91.0 \\ 
RSCnc & 085 & i &   30.0 &  275048824. &   57.9  &   41.5 \\ 
RSCnc & 085 & e &   30.0 &  275068331. &  153.4  &  237.1 \\ 
gamCru & 086 & i &  -62.3 &  275497768. &  182.2  &  217.3 \\ 
RSCnc & 087 & i &   30.0 &  276322450. &   58.8  &  105.9 \\ 
RSCnc & 087 & e &   30.0 &  276341685. &  152.5  &  303.7 \\ 
gamCru & 089 & i &  -62.3 &  277402848. &  181.9  &  359.0 \\ 
gamCru & 093 & i &  -62.3 &  280038476. &  203.3  &  102.7 \\ 
gamCru & 094 & i &  -62.3 &  280675202. &  192.0  &  135.7 \\ 
epsMus & 094 & i &  -72.8 &  280700017. &  257.0  &  286.5 \\ 
epsMus & 094 & e &  -72.8 &  280717828. &  316.7  &  136.4 \\ 
gamCru & 096 & i &  -62.3 &  282008576. &  187.6  &   56.4 \\ 
gamCru & 097 & i &  -62.3 &  282697254. &  187.5  &   11.4 \\ 
gamCru & 100 & i &  -62.3 &  285027485. &  213.1  &  169.4 \\ 
gamCru & 101 & i &  -62.3 &  285854639. &  213.1  &   37.1 \\ 
gamCru & 102 & i &  -62.3 &  286679833. &  212.9  &  281.4 \\ 
betPeg & 104 & i &   31.7 &  288910932. &  353.7  &  195.0 \\ 
RCas & 106 & i &   56.0 &  291029840. &   50.6  &  334.3 \\ 
betPeg & 108 & i &   31.7 &  292213793. &    4.3  &   74.5 \\ 
alpAur & 110 & i &   50.9 &  295141170. &  317.1  &  239.1 \\ 
alpAur & 110 & e &   50.9 &  295160215. &  241.7  &   78.6 \\ 
alpSco & 115 & i &  -32.2 &  302010739. &  170.6  &  294.1 \\ 
betPeg & 170 & e &   31.7 &  397982124. &   87.5  &  251.9 \\ 
betPeg & 172 & i &   31.7 &  401611701. &  302.8  &  257.5 \\ 
lamVel & 173 & i &  -43.8 &  403829302. &  131.2  &  284.9 \\ 
RLyr & 176 & i &   40.8 &  407915130. &  219.9  &   44.6 \\ 
RLyr & 176 & e &   40.8 &  407945440. &  164.1  &  149.1 \\ 
WHya & 179 & i &  -34.6 &  411898308. &  130.1  &  309.6 \\ 
RLyr & 180 & i &   40.8 &  412507880. &  217.2  &  211.4 \\ 
RLyr & 180 & e &   40.8 &  412535706. &  166.4  &  336.9 \\ 
WHya & 180 & i &  -34.6 &  413047699. &  130.4  &  341.2 \\ 
WHya & 181 & i &  -34.6 &  414196938. &  130.4  &   14.0 \\ 
RHya & 185 & i &  -29.4 &  418337590. &   45.1  &   31.6 \\ 
RHya & 185 & e &  -29.4 &  418351569. &  329.7  &  273.8 \\ 
RCas & 185 & i &   56.0 &  418059013. &  313.3  &  219.8 \\ 
muCep & 185 & e &   59.9 &  418027873. &   67.6  &  122.3 \\ 
\hline
\end{tabular}
\begin{tabular}{|c|c|c|c||c|c|c|} \hline
Star & Rev$^a$ & i/e$^b$ & B$^c$ &  ET$^d$ & Longitude$^e$ & m=0 Phase$^f$ \\ 
& & & (degrees) & (seconds) & (degrees) & (degrees) \\ \hline
WHya & 186 & e &  -34.6 &  419161471. &  292.1  &  286.9 \\ 
RDor & 186 & i &  -56.3 &  419057192. &  161.2  &   85.9 \\ 
gamCru & 187 & i &  -62.4 &  419913820. &  130.8  &   65.2 \\ 
gamCru & 187 & e &  -62.4 &  419936428. &  246.4  &  234.7 \\ 
RDor & 188 & i &  -56.3 &  420711012. &  160.7  &  185.6 \\ 
RDor & 188 & e &  -56.3 &  420716266. &  195.7  &  141.3 \\ 
WHya & 189 & e &  -34.6 &  421642104. &  291.2  &  257.2 \\ 
RCar & 191 & i &  -63.5 &  423385180. &  126.8  &  324.5 \\ 
RCas & 191 & i &   56.0 &  423126448. &  291.5  &  345.4 \\ 
muCep & 191 & i &   59.9 &  423048531. &  286.9  &  282.1 \\ 
muCep & 193 & i &   59.9 &  425115241. &  287.0  &  141.4 \\ 
RCas & 194 & e &   56.0 &  426266707. &   89.6  &  155.5 \\ 
2Cen & 194 & e &  -40.7 &  426591441. &  249.1  &  298.2 \\ 
muCep & 195 & i &   59.9 &  427182862. &  335.0  &  353.0 \\ 
WHya & 196 & i &  -34.6 &  430037069. &  157.8  &   54.3 \\ 
WHya & 196 & e &  -34.6 &  430051310. &  228.2  &  294.3 \\ 
WHya & 197 & i &  -34.6 &  432104346. &  158.7  &  268.8 \\ 
WHya & 197 & e &  -34.6 &  432118271. &  227.3  &  151.5 \\ 
RLyr & 198 & i &   40.8 &  435185387. &  267.4  &  218.1 \\ 
L2Pup & 199 & e &  -41.9 &  438488221. &  318.4  &   97.9 \\ 
RLyr & 199 & i &   40.8 &  439304272. &  248.7  &   59.2 \\ 
RLyr & 199 & e &   40.8 &  439331815. &  132.9  &  187.0 \\ 
RLyr & 200 & i &   40.8 &  442043470. &  264.0  &    9.9 \\ 
L2Pup & 201 & i &  -41.9 &  446129569. &   96.3  &  127.4 \\ 
RLyr & 202 & i &   40.8 &  447535799. &  303.7  &  154.0 \\ 
RLyr & 202 & e &   40.8 &  447559049. &   70.3  &  318.0 \\ 
lamVel & 203 & i &  -43.8 &  449043024. &   40.0  &   49.3 \\ 
lamVel & 203 & e &  -43.8 &  449096026. &  336.7  &  322.5 \\ 
L2Pup & 205 & i &  -41.9 &  456837673. &  130.3  &  226.6 \\ 
RLyr & 206 & i &   40.8 &  458836881. &  294.6  &  294.9 \\ 
RLyr & 208 & e &   40.8 &  464399829. &   64.3  &  203.7 \\ 
WHya & 236 & i &  -34.6 &  517599124. &  116.5  &  336.1 \\ 
2Cen & 237 & i &  -40.7 &  519786996. &   96.4  &  254.1 \\ 
2Cen & 237 & e &  -40.7 &  519835092. &   28.8  &  208.7 \\ 
alpSco & 237 & i &  -32.2 &  520137664. &  228.1  &  178.2 \\ 
alpSco & 237 & e &  -32.2 &  520153854. &  270.7  &   41.8 \\ 
betPeg & 237 & i &   31.7 &  520424269. &  265.5  &  282.4 \\ 
betPeg & 237 & e &   31.7 &  520430571. &  231.7  &  229.3 \\ 
alpSco & 238 & i &  -32.2 &  522206348. &  229.6  &   20.9 \\ 
alpSco & 238 & e &  -32.2 &  522221476. &  269.2  &  253.4 \\ 
alpSco & 239 & i &  -32.2 &  523688965. &  135.7  &  123.6 \\ 
RCas & 239 & i &   56.0 &  523881209. &   29.9  &  303.1 \\ 
RCas & 239 & e &   56.0 &  523893640. &   91.9  &  198.3 \\ 
rhoPer & 239 & e &   45.3 &  523938719. &  148.2  &  178.4 \\ 
alpSco & 241 & i &  -32.2 &  525831418. &  112.4  &   64.4 \\ 
alpSco & 241 & e &  -32.2 &  525849007. &    2.4  &  276.2 \\ 
alpSco & 243 & e &  -32.2 &  527916676. &    1.8  &  127.4 \\ 
RCas & 243 & i &   56.0 &  528004792. &  351.2  &  104.6 \\ 
alpSco & 245 & i &  -32.2 &  529597290. &  108.3  &    1.1 \\ 
alpSco & 245 & e &  -32.2 &  529610004. &  359.1  &  254.0 \\ 
gamCru & 245 & e &  -62.4 &  529546263. &  296.8  &   71.3 \\ 
\hline
\end{tabular}}}

\bigskip 
$^a$ Cassini Orbit Around Saturn

$^b$ i=ingress portion of occultation, e=egress portion of occultation

$^c$ Ring opening angle to the star (positive numbers correspond to stars in Saturn's northern hemisphere)

$^d$ Ephemeris Time in seconds past J2000 (TDB) when the spacecraft observed the inner edge of the Barnard Gap. 

$^e$ Inertial longitude measured from the ascending node of Saturn's ringplane on the J2000 reference plane

$^f$ Expected phase of an $m=0$ wave launched from the Prometheus 5:4 resonance (i.e. with $\omega= 728.28^\circ$/day), measured relative to the epoch time 2008-001T12:00:00 (ET 252460865.184)
\vspace{2in}
\end{table*}

\subsection{Occultation Observations}
\label{obs}

This examination of the region around the Barnard Gap uses stellar occultation data obtained by the Visual and Infrared Mapping Spectrometer (VIMS) onboard the Cassini Spacecraft \citep{Brown04}. While in its standard operating mode VIMS obtains spatially resolved spectra of various objects in the Saturn system, this instrument can also operate in a mode where it repeatedly measures the spectrum of a star as the rings pass between it and the spacecraft. In this mode, a precise timestamp is appended to each spectrum to facilitate reconstruction of the observation geometry \citep{Brown04}.  

We compute both the radius and inertial longitude in the rings that the starlight passed through using a combination of  the timing information accompanying each brightness measurement and the appropriate SPICE kernels \citep{Acton96}. Note that the information encoded in these kernels has been determined to be accurate to within one kilometer, and fine-scale adjustments based on the positions of circular ring features enable these estimates to be refined to an accuracy of  $\sim$ 150 m \citep{French17}.

Table~\ref{occs} lists the 101 occultation observations that we used in this analysis. This includes essentially all the occultations obtained prior to 2017 that covered the relevant part of the Cassini Division with adequate resolution and signal-to-noise to discern the quasi-periodic structures shown in Figure~\ref{baedge}. This large set of occultations spans an entire decade and includes observations obtained over a broad range of inertial longitudes, which ensures that we can uniquely identify the $m$-number and pattern speed of wave-like structures in the ring. However, for some aspects of this analysis we focus our attention on a sub-set of these observations: the occultations of the star $\gamma$ Crucis obtained between Cassini ``Revs" (that is, orbits) 71 and 102 in late 2008 and  early 2009. All of these occultations use the same star and occurred at about the same inertial longitude, and have very good signal-to-noise, all of which facilitates comparisons among these profiles.

For this particular study, we only consider data obtained at wavelengths between 2.87 and 3.00 microns, where the rings are especially dark and so provide a minimal background to the stellar signal. Together with the highly linear response of the instrument \citep{Brown04}, this low ring background means that the raw signal is directly proportional to the transmission $T$ through the rings. In practice, the transmission is estimated by normalizing the signal to unity during a time when the star was not obscured by the rings. Whenever possible, the selected time period corresponds to when the star was visible through the Huygens Gap in the Cassini Division (i.e. 117,700-117,750 km from Saturn center), and if this region was not available, a time period when the star was outside the main rings (i.e. more than 145,000 km from Saturn's center) was used. Also, any instrumental background level was removed by subtracting a constant offset from the data equal to the mean signal level when the star was behind an opaque part of the B ring (105,700-106,100 km). The resulting transmission values $T$ can then be transformed into the ring's normal optical depth $\tau_n$ using the standard expression $\tau_n=-\ln(T)\sin|B|$, where $B$ is the elevation angle of the star above the ring (see Table~\ref{occs}).  Note that for low optical depth regions like the Cassini Division, $\tau_n$ should be largely independent of the observation geometry and directly proportional to the surface mass density $\sigma$.

\subsection{Wavelet analysis methods}
\label{wave}

We analyze these occultation data using wavelet-based tools developed in \citet{HN16} for isolating wave signatures in Saturn's B ring. These tools are designed to take multiple occultation profiles and combine the data in a manner that isolates signals with pattern speeds and $m$-values consistent with specified density waves. Details of this approach are provided in \citet{HN16}, but for the sake of completeness we will summarize the basic method here.

We begin by taking each occultation profile, interpolating the transmission estimates onto a regular grid of radii with a spacing of 100 meters, converting the transmission values to normal optical depth,\footnote{Note that \citet{HN16} applied the wavelet transformation directly to the transmission profiles instead of the optical depth profiles. This was a sensible choice because that work only considered occultations obtained with similar geometries and ring regions with high optical depth. However, in this case we are considering a broader range of occultation geometries and a ring region with low optical depth. Converting the profiles to optical depth helps make the signals observed at different times more comparable and simplifies the interpretation of the reconstructed wave profiles.}  and transforming the profile into a wavelet  using the IDL  {\tt wavelet} routine  \citep{TC98} with a Morlet mother wavelet and parameter $\omega_0=6$.This yields a complex two-dimensional wavelet for each profile $\mathcal{W}_i=\mathcal{A}_ie^{i\Phi_i}$ where $\mathcal{W}_i, \mathcal{A}_i$ and $\Phi_i$ are all functions of both radius $r$ and radial wavenumber $k$. For the signal from a density wave the wavelet phase $\Phi_i$ is equivalent to the local wave phase $\phi_r+\phi_{\lambda t}$. Given the observed longitude $\lambda_i$ and observation time $t_i$ for each occultation, we can compute the following phase parameter

\begin{equation}
\phi_i=|m|\lambda_i-\omega (t_i-t_0)
\end{equation} 
where $|m|$ and $\omega$ are the assumed $m$-number and frequency of the perturbation, and $t_0$ is an arbitrary epoch time. Here $t_0$ corresponds to 2008-001T12:00:00 UTC, in order to be consistent with the epoch time used by \citet{French16}. For a wave with the selected $m$ and $\omega$ values, the phase difference $\Phi_i-\phi_i$ will be the same function of radius $r$ ($\phi_r$) for every occultation, and so we can define a phase-corrected wavelet:

\begin{equation}
\mathcal{W}_{\phi,i}(r,k)=\mathcal{W}_i(r,k)e^{-i\phi_i}=\mathcal{A}_i(r,k)e^{i(\Phi_i(r,k)-\phi_i)}
\end{equation} 
For any signal with the selected values of $m$ and $\omega$, the phase of this corrected wavelet should be the same for all the occultation profiles. The average phase corrected wavelet of $N$ profiles:

\begin{equation}
\langle \mathcal{W}_\phi(r,k) \rangle=\frac{1}{N}\sum_{i=1}^N\mathcal{W}_{\phi,i}(r,k)
\end{equation}
will therefore be nonzero for such patterns while any other structure will average to zero.  Only patterns with the desired $m$ and $\omega$ should therefore remain in the power of the average phase corrected wavelet

\begin{equation}
\mathcal{P}_\phi(r,k)=|\langle \mathcal{W}_\phi(r,k) \rangle|^2=\left|\frac{1}{N}\sum_{i=1}^N\mathcal{W}_{\phi,i}(r,k)\right|^2
\end{equation}
while all other signals are only seen in the average wavelet power:

\begin{equation}
\bar{\mathcal{P}}(r,k)=\langle |\mathcal{W}_\phi(r,k)|^2 \rangle=\frac{1}{N}\sum_{i=1}^N\left|\mathcal{W}_{\phi,i}(r,k)\right|^2
\end{equation}
We also use the ratio of these two powers $\mathcal{R}(r,k)=\mathcal{P}_\phi/\bar{\mathcal{P}}$, which ranges between 0 and 1 \citep{HN16}, as a measure of how much of the signal at a given $r$ and $k$ is consistent with the assumed $m$  and $\omega$.

\subsection{Results}
\label{results}

\begin{figure}
\centerline{\resizebox{3.5in}{!}{\includegraphics{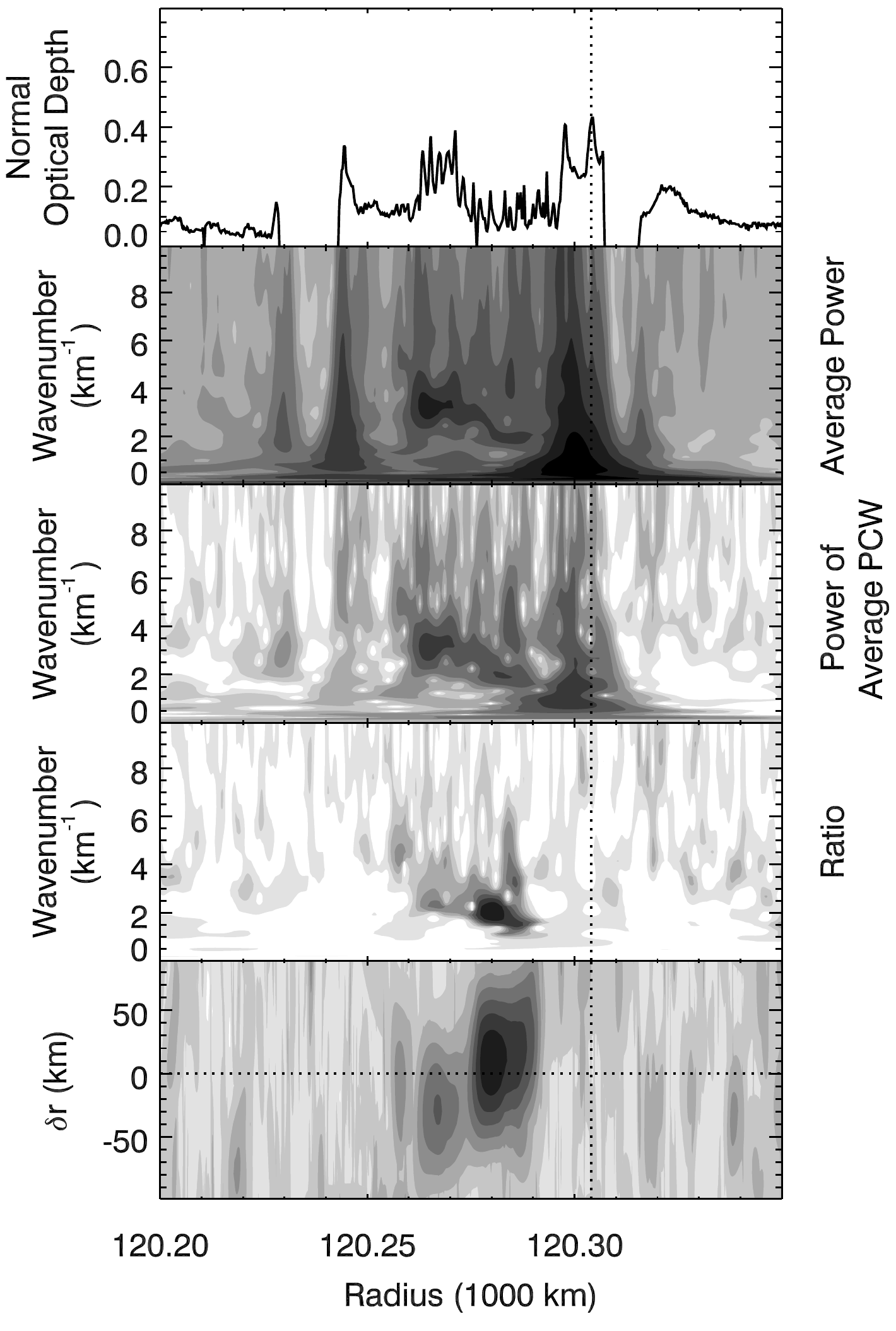}}}
\caption{Wavelet analysis of the structures interior to the Barnard gap using data from $\gamma$-Crucis occultations obtained in 2008-2009. The top panel shows a representative occultation profile for reference. The second panel shows the average wavelet power, which contains strong signals near the various sharp edges, as well as a suggestive diagonal band between 120,260 and 120,290 km. The third panel shows the power of the average phase-corrected wavelet assuming $m=0$ and a pattern speed of 728.28$^\circ$/day, which corresponds to the radial epicyclic frequency at the location of the Prometheus 5:4 resonance (i.e. 120304 km from Saturn center, marked by a dotted vertical line). The fourth panel shows the power ratio, which only contains the signal from the region interior to the Barnard gap. Note that the second and third panels use a common set of logarithmically-spaced greyscale levels, while the ratio uses linearly-spaced levels between 0 and 1. The bottom panel shows the peak power ratio as a function of radius and assumed epicyclic frequency (expressed in terms of a shift in the resonant radius), which shows that the signal is strong only when the pattern speed is close to the assumed value (marked with a horizontal dotted line). }
\label{wavegam}
\end{figure}

We used the above tools to search for patterns in the region between the Bessel and Barnard Gaps with values of $m$ between $-10$ and 10 and, for each $m$, a range of pattern speeds corresponding to the expected values of $\omega$ within 100 km of the Barnard gap inner edge (cf. Equation~\ref{eqpat}). The only strong signal found with this approach was obtained with $m=0$, corresponding to an axisymmetric wave with a pattern speed $\Omega_p=\omega$ equivalent to the local epicyclic rate. Figure~\ref{wavegam} shows the signal in the $\gamma$ Crucis occultations obtained in 2008 and 2009. Both the  power of the average phase-corrected wavelet and the power ratio show a clear diagonal band  running from 120,260 km to 120,290 km, consistent with an inward-propagating density wave launched from somewhere close to the inner edge of the Barnard gap. Also, the signal in the power ratio is strongest when we assume the appropriate pattern speed for an $m=0$ wave launched from that location, providing further evidence that this region does indeed contain an $m=0$ wave (but see the Appendix for a potential ambiguity between $m=0$ density waves and $m=2$ bending waves).

\begin{figure}
\centerline{\resizebox{3.5in}{!}{\includegraphics{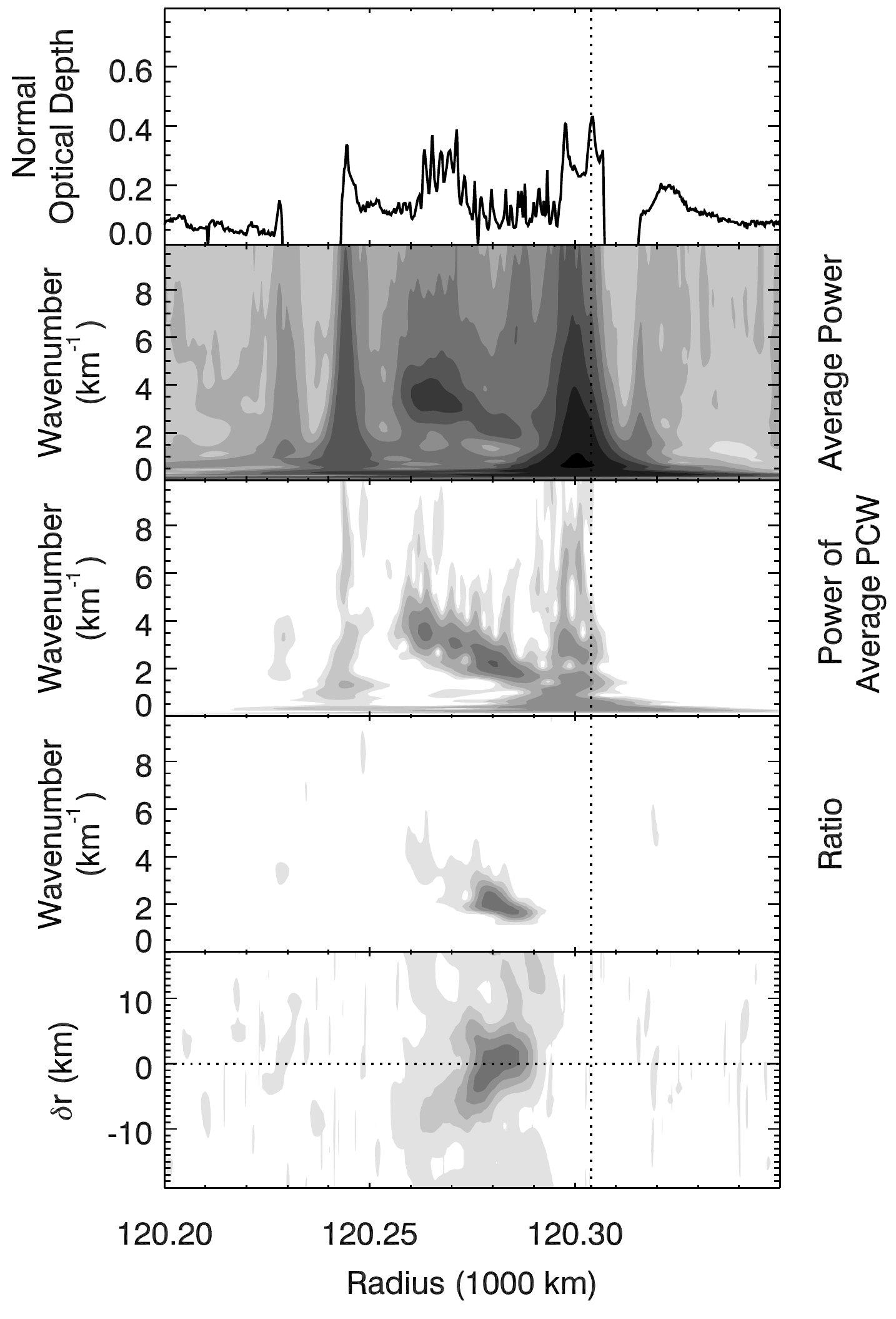}}}
\caption{Wavelet analysis of the structures interior to the Barnard gap using the full set of occultations. See Figure~\ref{wavegam} for detailed descriptions of the panels. Note that  the vertical range of the bottom plot is expanded in this case to better show the signal. There is no peak interior to 120,275 km outside the illustrated range.}
\label{wavefull}
\end{figure}

Figure~\ref{wavefull} shows the same analysis for the full set of 101 occultations. For this full data set, the same wave signal can be seen in the average phase-corrected wavelet between 120,260 and 120,290 km, although the signal in the power ratio plot is quite a bit weaker than it is for the 2008-2009 $\gamma$ Crucis occultations (especially interior to 120,275 km, see below).  Even so, it is still the case that no other $m$ value yields a sensible signal for this structure. Moreover, this full data set removes the ambiguity between $m=0$ density waves and $m=2$ bending waves that exists when only the $\gamma$ Crucis occultations are considered (see Appendix).   Furthermore, the extended data set yields a much tighter constraint on the wave's pattern speed, which for radii between 120,275 and 120,290 km is 728.28$\pm$0.02$^\circ$/day. This narrow range of pattern speeds implies that the resonance responsible for generating this wave lies at 120304.0 +/- 2 km, which includes the nominal location of the Prometheus 5:4 resonance at 120304.0 km and 
the mean position of the Barnard gap inner edge at 120303.7 km (French {\it et al.} 2016a). This location is also consistent with the observed trends in the pattern's wavenumber (that is, $k$ must approach zero at the exact resonance).

\begin{figure}
\resizebox{3.5in}{!}{\includegraphics{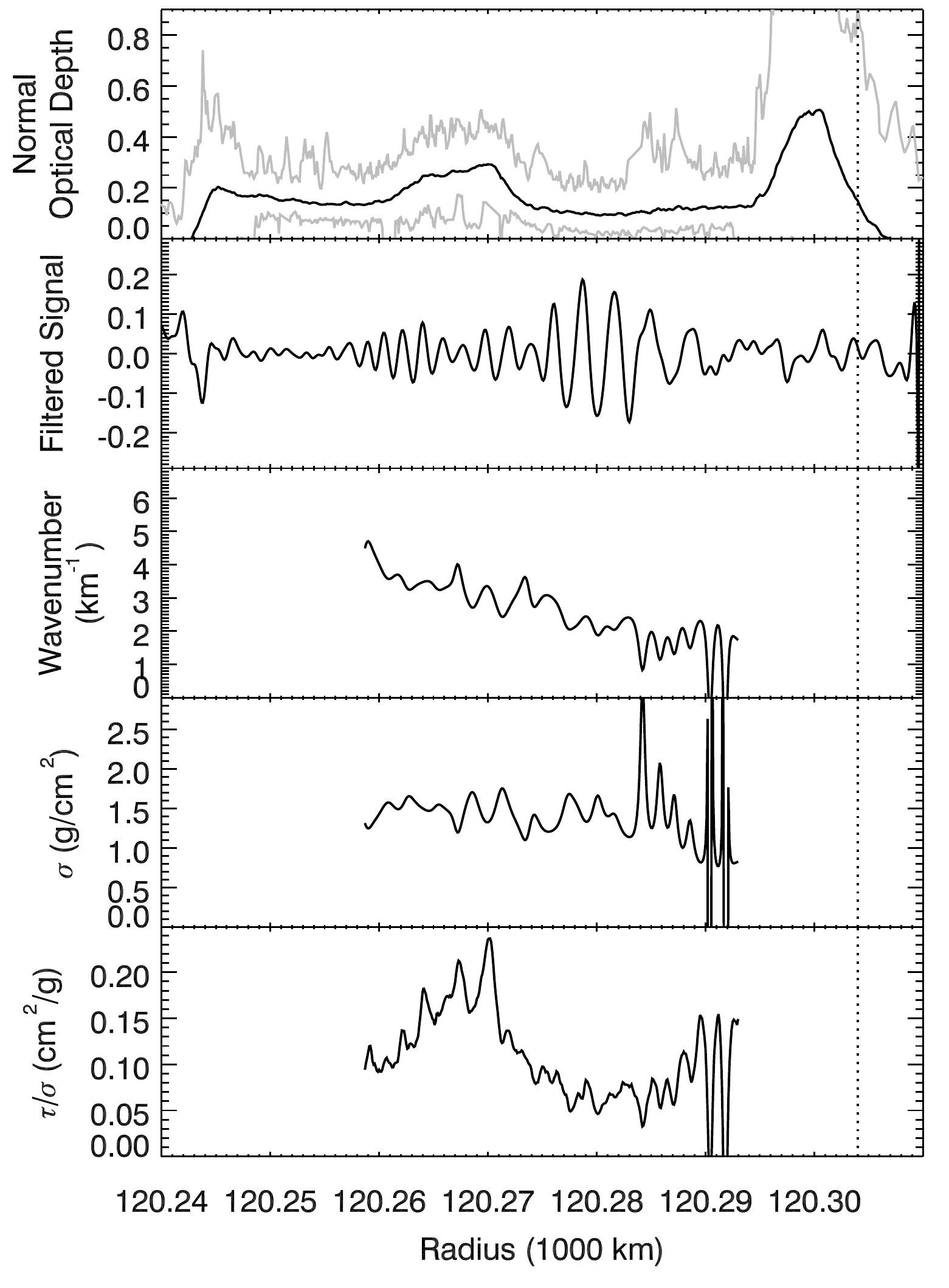}}
\caption{Reconstructed wave profile based on the full set of all occultations used to make Figure~\ref{wavefull}. The top panel shows the mean and range of normal optical depths among the profiles. The second panel shows the reconstructed profile of the fractional optical depth variations for the $m=0$ wave extracted from the average phase-corrected wavelet assuming a pattern speed of 728.28$^\circ$/day, corresponding to a resonant radius of 120304 km (marked by the dotted line).   The third profile shows the radial  wavenumber of this wave, while the fourth and fifth panels show the estimated surface mass density and opacity derived from this wave (again, assuming a fixed resonant radius of 120304 km). The latter two values are consistent with other estimates of these parameters for comparable regions of the Cassini Division.}
\label{profmod}
\end{figure}

In order to verify that this wave-like signal is indeed an axisymmetric density wave, we can take the average phase-corrected wavelet $\langle \mathcal{W}_\phi \rangle$ and apply the inverse wavelet transform to obtain a reconstructed profile of the $m=0$ signal. Note that this profile is itself complex, but the real and imaginary parts are simply the wave profiles at times when $\phi_{\lambda t}$ = 0 and $\pi/2$, respectively. Figure~\ref{profmod} shows the reconstructed profile derived from the full set of occultation profiles.\footnote{Reconstructed profiles generated with subsets of the data like that shown in Figure~\ref{wavegam} have a generally similar structure. However, the wavelength trends are noisier because residual background noise from other structures is not as cleanly removed in the average phase-corrected wavelet when fewer occultations are considered.} This profile was computed using only wavenumbers between $2\pi/1$ and $2\pi/10$ km$^{-1}$ in order to filter out slow variations and high-frequency noise. The resulting profile indeed  looks like an inward-propagating density wave, with a wavelength that steadily decreases with distance from the putative resonance.

To better quantify the trends in this feature's wavelength, we compute the phase $\phi_r$ from the real and imaginary parts of the profile and then take the radial derivative of this quantity to determine the radial wavenumber of the signal $k$. This parameter shows a linear trend, consistent with Equation~\ref{keq} for a density wave in a region of near-uniform mass density.  We can then use this equation, assuming $m=0$ and $r_L=120,304$ km, to estimate both the surface mass density $\sigma$ and the opacity parameter $\tau_n/\sigma$.  We find that over the region covered by the wave the surface mass density varies between 1.0 and 1.5 g/cm$^2$. This surface mass density is comparable to estimates derived from the Prometheus 9:7, Pan 6:5 and Atlas 5:4 waves in the inner Cassini Division, which all fall between 1.1 and 1.4   g/cm$^2$  and occur in regions of comparable optical depth \citep{Colwell09}. These consistent numbers provide further evidence that this structure is indeed an $m=0$ density wave. 

Intriguingly, the surface mass density remains nearly constant despite the optical depth having a clear peak around 120,270 km, which means the opacity $\tau_n/\sigma$ varies substantially across this region. This behavior is actually consistent with previous analyses of the A, B and C rings, which show that the surface mass density is far less variable than the optical depth \citep{Tiscareno13, HN14, HN16}. 


While the wavelength trends associated with this structure are perfectly consistent with those expected for a $m=0$ density wave, closer inspection of the wavelets and the reconstructed wave profile in Figures~\ref{wavegam}-\ref{profmod} reveal some surprising variations in the pattern's amplitude and pattern speed. The average wavelet power $\bar{\mathcal{P}}$ for both the $\gamma$ Crucis occultations and the full data set shows the strongest wave signal interior to 120,275 km. By contrast, for both figures the power of the average phase-corrected wavelet ${\mathcal{P}}_\phi$ has a more uniform strength  between 120,260 and 120,290 km. In fact, for the full data set (Figure~\ref{wavefull}) the signal in ${\mathcal{P}}_\phi$ is higher around 120,280 km than it is around 120,265 km, which is consistent with the reconstructed profile (Figure~\ref{profmod}) having a higher amplitude around 120,280 km than around 120,265 km. Finally, the signal in the power ratio $\mathcal{R}$ is strongest between 120,275 and 120,290 km, and is much weaker interior to 120,275 km. This trend in $\mathcal{R}$ is consistent with the above trends in $\bar{\mathcal{P}}$ and $\mathcal{P}_\phi$, and implies that the signals exterior to 120,275 km are more consistent with those expected for an axisymmetric density wave with the given pattern speed.  

The plots of the power ratio versus radius and pattern speed reveal another important difference between the inner and outer parts of this wave. In both Figures~\ref{wavegam} and~\ref{wavefull}, the peak signal exterior to 120,275 km occurs at a pattern speed consistent with that expected for an $m=0$ wave launched from the inner edge of the Barnard gap. By contrast, interior to 120,275 km the peak signal seems to shift to smaller resonant radii (that is, higher pattern speeds). For the $\gamma$ Crucis data shown in Figure~\ref{wavegam}, the best-fit pattern speed for this part of the wave would correspond to a resonant location 20 km interior to the gap edge. Other sub-sets of the data show a similar general trend, but the best-fit resonant radius varies by several kilometers. For the full data set the peak power ratio is considerably weaker than that found outside 120,275 km, and the best fit resonant radius is only 8 km interior to the gap edge. 

\begin{figure}
\resizebox{3.4in}{!}{\includegraphics{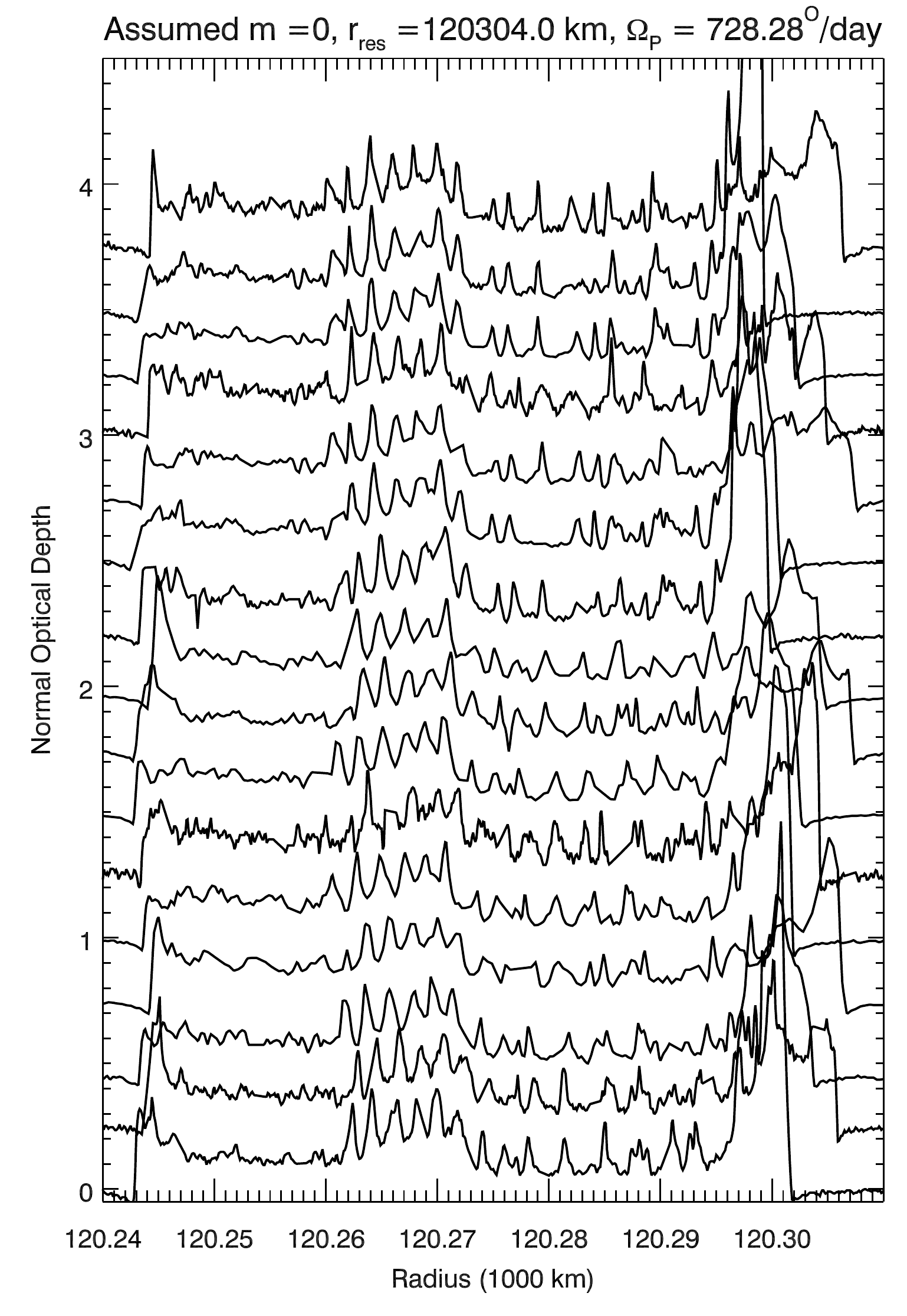}}
\caption{Selected $\gamma$ Crucis occultation profiles sorted by phase of the $m=0$ pattern (phase increases upwards). Note that the position of many peaks shift from left to right as one moves from top to bottom of this plot, consistent with an $m=0$ wave.}
\label{prof1}
\end{figure}

\begin{figure}
\resizebox{3.4in}{!}{\includegraphics{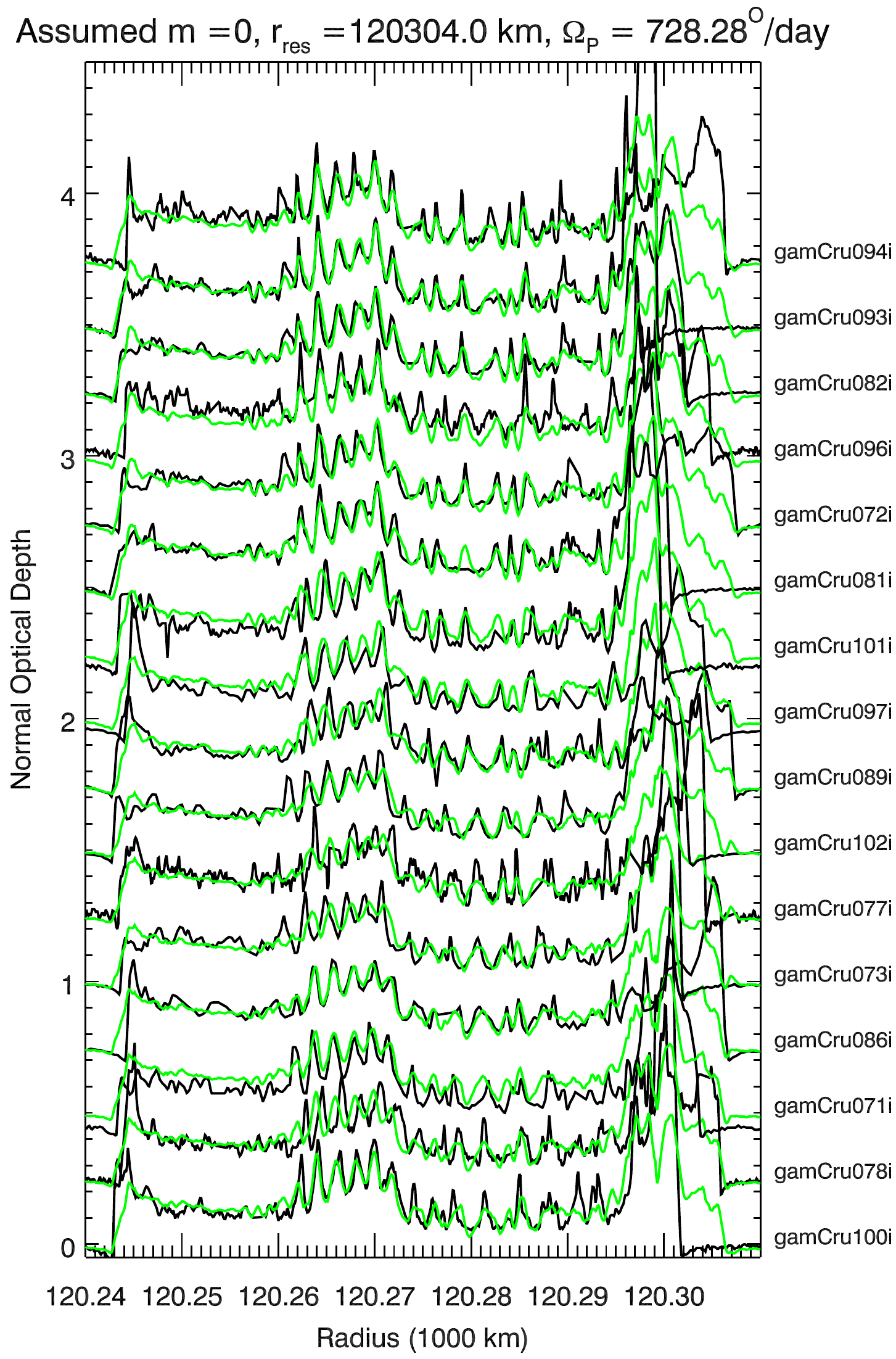}}
\caption{Selected $\gamma$ Crucis occultation profiles sorted by phase of the $m=0$ pattern (phase increases upwards). The overlaid green curves show the predicted variations expected to arise from the $m=0$ wave based on the average phase-corrected wavelet and the appropriate phase factors $\phi_i$ (these are scaled by a factor of 2 for clarity). Note that interior to 120,275 km there are noticeable mismatches between the expected and observed profiles. Such misalignments are less common at larger radii.}
\label{prof1p}
\end{figure}

Together, these trends imply that while the periodic optical depth variations are more prominent interior to 120,275 km,  the signals exterior to 120,275 km are more coherently organized. Examinations of the wave profiles from individual occultations confirm these findings. Figure~\ref{prof1}  shows  occultation profiles derived from the 2008-2009 $\gamma$ Crucis occultations sorted by the phase $\phi_{\lambda t}=-\omega (t-t_0)$ for the $m=0$ wave with $\omega=728.28^\circ$/day. The comparable viewing geometries of these occultations facilitates comparisons among them. For all these profiles, the most obvious wave-like signals lie between 120,260 and 120,275 km. By contrast, between 120,275 and 120,290  km the situation is more complicated because the peaks are not strictly periodic. These aspects of the individual profiles primarily reflect the trends in the average wavelet power $\bar{\mathcal{P}}$, which also indicate a stronger periodic signal interior to 120,275 km. 

Assessing how consistent these structures are with an $m=0$ wave is not as straightforward. Since the profiles are sorted by phase, the positions of optical depth maxima and minima should shift from left to right as we proceed from the top to the bottom of the plot. The peaks interior to 120,275 km do  generally follow this trend, while the situation further out is less obvious. In order to clarify this situation, Figure~\ref{prof1p} shows the same occultation profiles together with profiles of what the predicted $m=0$ wave pattern for each observation should look like based on the average phase-corrected wavelet of these occultations (these predicted wave signals are superimposed on the average background ring profile to facilitate comparisons). While the variations associated with the wave are easier to see interior to 120,275 km, the locations of the individual peaks and troughs in this region deviate from those expected for the $m=0$ wave in several of the profiles. By contrast, exterior to 120,275 km the peaks are less obvious, but their positions are generally much better aligned with those expected for the $m=0$ wave. These comparisons therefore support the notion that while the wave amplitude is higher in the inner part of the wave, the outer part of the wave has a more coherent $m=0$ pattern.

\begin{figure}
\resizebox{3.4in}{!}{\includegraphics{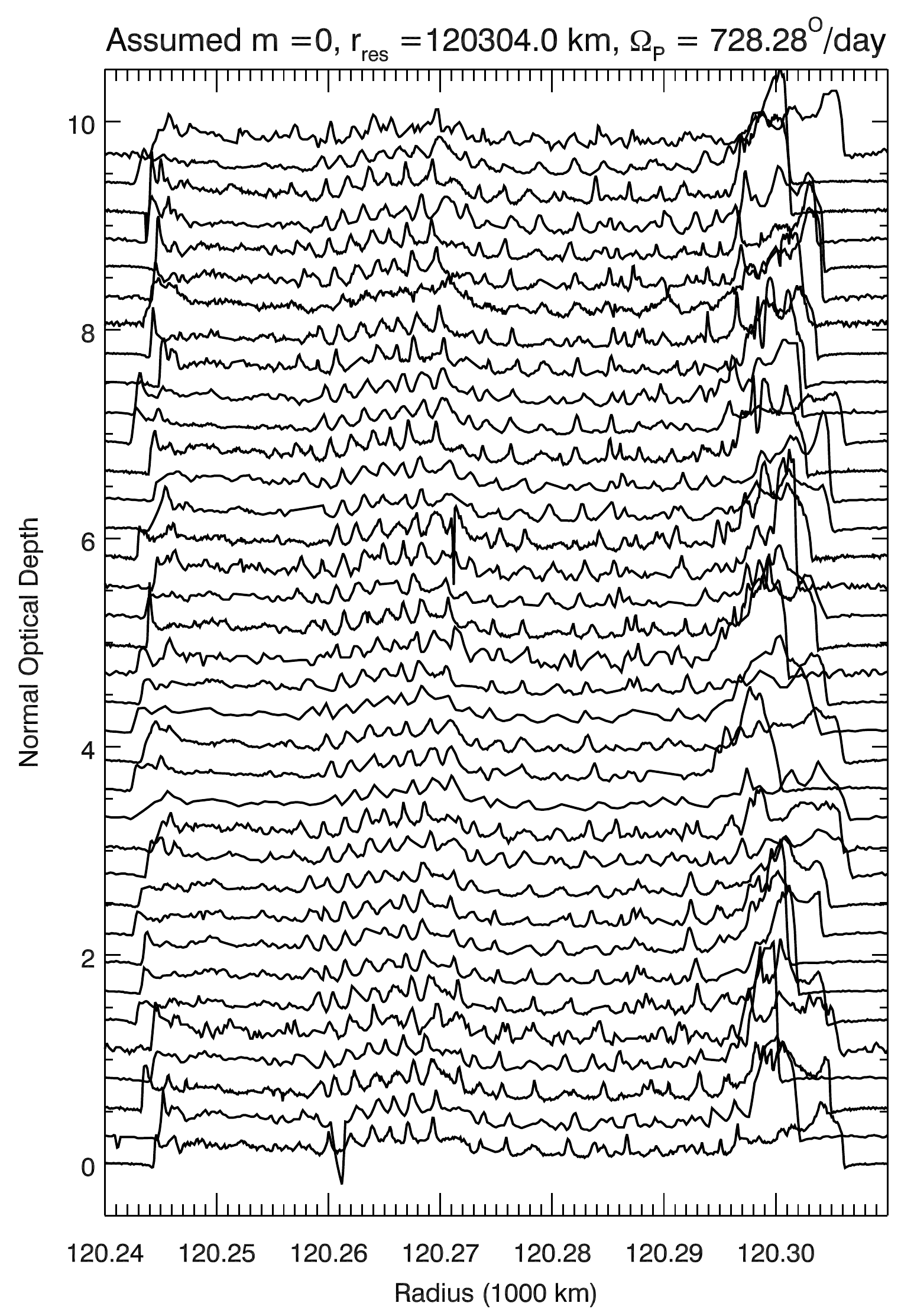}}
\caption{Occultation profiles from 2012-2014 sorted by phase of the $m=0$ pattern (phase increases upwards). Note that the locations of the prominent density variations around 120,265 km are shifted slightly inwards relative to those shown in Figure~\ref{prof1}.}
\label{prof2}
\end{figure}


Other subsets of the occultation data show similar differences between the inner and outer parts of this wave, and provide evidence for longer-term changes in the wave's structure.  Figure~\ref{prof2} shows profiles obtained between 2012 and 2014. Note that these occultations have more heterogeneous viewing geometries, which makes the signal-to-noise of the density variations more variable. Despite this, the radial extent of the strong wave pattern around 120,265 km in these profiles appears to differ systematically from that found in Figure~\ref{prof1}. In the earlier $\gamma$ Crucis occultations, the strong peaks are found in a region between 120,262 km and 120,272 km, with peaks rarely being seen interior to 120,260 km. However, for the data obtained between 2012 and 2014  the strong peaks are often found around 120,260 km and now are rarely found exterior to 120,270 km. This packet of waves therefore appears to have shifted roughly 2 km inwards during the 4 years between these observations.

The observed inward shift of this wave packet is consistent with the expected inward group velocity for this density wave. Assuming a surface mass density of around 1.2 g/cm$^2$, Equation~\ref{group} yields a group velocity of roughly 0.5 km/year, which would be consistent with a shift of around 2 km in the 4 years between the two groups of observations. We may therefore posit that the amplitude of the perturbation driving the wave changed over time, and a packet of high-amplitude waves created sometime in the past (ending around 70 years ago) has propagated inwards over the course of the Cassini mission. 

Time variations in how this wave is excited  could also help explain why the wave becomes less well organized further away from the nominal resonance. If not only the perturbation's amplitude, but also its frequency changed over time, then this wave may have some similarities with the waves excited by resonances with the satellite Janus, whose orbit period changes periodically due to interactions with its coorbital companion Epimetheus \citep{Yoder83}. The waves generated by Janus  exhibit  unusual variations in their pattern speeds that likely arise because parts of the wave have partially decoupled from the satellite perturbations \citep{Porco05, Tiscareno06, HN16}, and a similar phenomenon could be operating here. In principle, these variations in the wave's pattern speed could   impact the estimates of the ring's surface mass density and opacity shown in Figure~\ref{profmod}. However, since there is not yet a complete theory for how waves with variable pattern speeds propagate in shearing disks, there are not yet any quantitative estimates for how big these effects may be. Developing such a theory is well beyond the scope of this article, but we can explore this idea further by considering how this axisymmetric wave could be excited in the first place.

 \section{Excitation of axisymmetric density waves from resonantly-confined edges}
\label{model}

The above observations provide reasonably firm evidence that an axisymmetric $m=0$ wave is being  launched from the inner edge of the Barnard gap. Hence there must be a perturbation that is independent of longitude but oscillates in time at a frequency equal to the radial epicyclic frequency of the particles within two kilometers of the gap edge. In principle, such a perturbation could arise from collisions or the gravitational potential, but for the sake of simplicity we will assume here that it is due to a term in the gravitational potential that can be written as:

\begin{equation}
\mathcal{U}_0=C_0e^{\pm i\kappa_L t}
\end{equation} 
where $\kappa_L$ is the radial epicyclic frequency at the gap edge and $C_0$ is a constant that determines the strength of this wave. 


The fact that the wave is launched from an edge that is  perturbed by the 5:4 resonance with the satellite Prometheus strongly suggests that the perturbations from that satellite play a role in generating this wave. We may designate the term in the potential associated with this resonance as:

\begin{equation}
\mathcal{U}_P
=C_Pe^{i5(\lambda-n_Pt)} =C_Pe^{i(5\lambda-(5 n_L-\kappa_L) t)}
\end{equation} 
where  $n_P$ is the mean motion of  Prometheus, and the last equality uses the definition of the resonant mean motion $n_L$  and radial epicyclic frequency $\kappa_L$ (see Equation~\ref{eqpat}). The constant $C_P$ quantifies the strength of the perturbation from this moon. 


Of course, the perturbation from Prometheus does not have the same form as $\mathcal{U}_0$, 
but \citet{French16} determined that the inner edge of the Barnard gap also exhibits a diverse array of normal modes (see Table~\ref{modes}). Each of these normal modes has an azimuthal wavenumber $m$ and a pattern speed $\Omega_p=n_L-\kappa_L/m$. The only detectable normal modes have pattern speeds slower than the local orbital rate (i.e. positive values of $m$). While we do not yet have a complete understanding of what determines the relative amplitudes of these normal modes, it is likely that the observable variations in the edge position correspond to cavity modes trapped between the edge and a nearby location within the ring where the resonant condition is exactly satisfied. For material close to the gap edge, the variations in the edge position should generate terms in the local gravitational potential that have $m-$fold symmetry and rotate at the corresponding pattern speed:

\begin{equation}
\mathcal{U}_m =C_m e^{i(m\lambda-(mn_L-\kappa_L)t)}.
\end{equation}
Each of these modes could potentially produce azimuthal variations in the local gravitational field or perturb the motions of particles via interparticle collisions. Since the latter is a dissipative process, it is not strictly appropriate to represent this force in terms of a potential.  Future studies should examine this more carefully, but for the sake of simplicity we will assume here that the perturbation associated with each edge mode can be written in the above form.

\begin{table}
\caption{Modes observed on the inner edge of the Barnard Gap \citep{French16}}
\label{modes}
{\begin{tabular}{|c c c|}\hline
Mode & Amplitude (km) & Phase et Epoch$^a$\\ \hline
1 & 0.44$\pm$0.06 & 200.07$\pm$8.92 \\
2$^b$ & 0.61$\pm$0.07 & 44.12$\pm$3.25 \\
3 & 1.31$\pm$0.06 & 108.47$\pm$ 1.02 \\
4 & 1.64$\pm$0.07 & 46.00$\pm$ 0.57 \\
5 & 1.36$\pm$0.06 & 27.61$\pm$ 0.56$^c$ \\
6 & 0.59$\pm$0.07 & 24.56$\pm$ 0.57 \\
7 & 0.55$\pm$0.06 & 46.93$\pm$ 0.95 \\
8 & 0.30$\pm$0.07 & 10.41$\pm$ 1.59 \\
9 & 0.71$\pm$0.07 & 8.38$\pm$ 0.55 \\
10 & 0.42$\pm$0.06 & 1.75$\pm$ 0.98 \\
13 & 0.36$\pm$0.06 & 26.84$\pm$ 0.86 \\ \hline
\end{tabular}}

\bigskip

$^a$ Epoch is UTC 2008-001T12:00:00

$^b$ Free $m=2$ mode, note that the $m=2$ mode driven by the Mimas 2:1 resonance has an amplitude that is about three times smaller.

$^c$ Phase determined from the data, the expected phase of the $m=5$ resonance with Prometheus is 23$^\circ$.

\end{table}

It is important to note that \citet{French16} found no evidence for an $m=0$ mode in the position of the Barnard Gap's inner edge. This is probably because such a mode, like Outer Lindblad Resonances, produces disturbances that propagate inwards from the edge, and so cannot become a self-excited cavity mode trapped close to the edge. Instead, we posit that the $m=0$ wave is generated through a non-linear mixing of the above terms in the local gravitational potential. A detailed model of how this mixing could occur is beyond the scope of this paper, and so we instead simply demonstrate that appropriate mixtures of $\mathcal{U}_P$ and two different $\mathcal{U}_m$ can produce the desired source term for the $m=0$ wave.

Since each normal mode on the edge has a different prescribed dependence on longitude, there is no way to mix $\mathcal{U}_P$ with a single $\mathcal{U}_m$ to  produce a term that is independent of $\lambda$ like $\mathcal{U}_0$. However, if we consider a process that mixes $\mathcal{U}_P$ with two $\mathcal{U}_m$ terms, we can get terms that look like:

\begin{equation}
\mathcal{F}\mathcal{U}_P\mathcal{U}^*_m\mathcal{U}^*_{m'} = \mathcal{F}C_PC^*_mC^*_{m'}e^{i(5-m-m')(\lambda-n_Lt)-i\kappa_Lt},
\end{equation}
where asterisks denote complex conjugates, and $\mathcal{F}$ is a factor that describes the efficiency of the interference. This expression will have the desired form so long as we chose $m+m'=5$. Similar terms can be obtained from combinations like $\mathcal{U}_P\mathcal{U}^*_m\mathcal{U}_{m'}, \mathcal{U}_P\mathcal{U}_m\mathcal{U}^*_{m'}$, etc.,\footnote{All these terms are  possible because the physical potential  (i.e. the real part of $\mathcal{U}$)  contains products of terms that go like $\sin[m\lambda-(mn_L-\kappa_L)t]$, and the expansion of these products into sum and difference frequencies yields terms equivalent to the real parts of all these terms.} which would yield suitable terms if $m=5+m'$ or $m'=5+m$. However, since the strongest normal modes observed on the edge have $m<5$ (see Table~\ref{modes}), we will focus on the first option here. Note that there are two different mode combinations that can produce the desired mixture, one with $m=1$ and $m'=4$ and another with $m=2$ and $m'=3$. Both of these pairs include one of the highest amplitude modes ( $>1$ km in  Table~\ref{modes}) and another mode with amplitude around 0.5 km.

Of course, without an actual theory for how these perturbations actually interact with each other, we cannot make quantitative predictions for the amplitude or the phase of this perturbation to see if they are sufficient to produce the observed wave. However, we can at least determine what the amplitude  of these perturbations would need to be in order to be consistent with the observed wave signature. 

The fractional optical depth variations associated with the axisymmetric wave are around 50\% (see Figure~\ref{baedge} and Figure~\ref{profmod}), which is comparable to the variations seen in nearby density waves generated by Pan. This implies that the ratio $\mathcal{U}_0/\mathcal{U}_{P}$ should be of order the mass ratio between Pan and Prometheus, which is about 3\% \citep{Porco07, Jacobson08}. Hence we can conclude that in order for the edge modes to produce the observed waves  $|\mathcal{F}\mathcal{U}^*_m\mathcal{U}^*_{m'}| \sim 0.03$. This is not many orders of magnitude less than one and so implies that the mixing between the modes needs to be fairly strong.

Even if the  mixing between the relevant modes is reasonably strong, the axisymmetric terms arising from different combinations of terms could potentially interfere with each other. Fortunately, we can estimate the relative phases of the axisymmetric terms by re-writing them in the following form:

\begin{equation}
\mathcal{F}\mathcal{U}_P\mathcal{U}^*_m\mathcal{U}^*_{m'} = \mathcal{F}C_PC^*_mC^*_{m'}e^{-i(5\lambda_P-m\lambda_m-m'\lambda_{m'})}
\end{equation}
where $\lambda_P$ is the longitude of Prometheus, while $\lambda_m$ and $\lambda_{m'}$ are longitudes that track the two normal modes. Since $\lambda_P$ corresponds to a minimum in the radial position of the Barnard Gap edge, $\lambda_m$ and $\lambda_{m'}$ should also correspond to minima of their corresponding edge modes. These numbers correspond to the phase parameters provided in Table~\ref{modes} from \citet{French16}. Hence $5\lambda_P-m\lambda_m-m'\lambda_{m'}$ is the analog of $m\lambda_{sat}$ for standard Lindblad resonances. Using the phases given in Table~\ref{modes}, we find that for the  $m=2/m'=3$ mode combination that this phase is 61$^\circ$-83$^\circ$, while for the $m=1/m'=4$ mode combination it is 91$^\circ$-116$^\circ$ (the lower numbers use the actual numbers for Prometheus' longitude, while the higher ones use the observed location of the $m=5$ edge mode at epoch). These phase parameters are very close to each other, indicating that these two perturbations are nearly in phase with each other. Thus these two different mode combinations could reinforce each other, supporting the formation of an $m=0$ wave. 

Finally, we can note that this basic scenario can potentially accommodate the unusual trends in the wave's amplitude and coherence. Note that the $m=5$ structure on the Barnard gap edge is not perfectly aligned with Prometheus (the phase of the pattern at epoch is about 5$^\circ$ away from its expected orientation relative to the moon). Similar offsets have been observed in the $m=2$ structure in the B-ring's outer edge, which occur because there are actually two separate $m=2$ patterns on this edge, a ``forced'' pattern that tracks  Mimas and a ``free'' pattern moving at a slightly different pattern speed \citep{SP09, Nicholson14b}. Together, these two patterns produce a  combined $m=2$ edge structure whose amplitude and orientation relative to Mimas varies slowly over time. A similar phenomenon could potentially occur on the Barnard gap inner edge, causing the amplitude and phase of the $m=5$ term to vary and thus producing variations in the wave amplitude that propagate inwards. Looking at the reconstructed wave profile in Figure~\ref{profmod}, we can posit that the peaks in amplitude around 120,280 km and 120,265 km could represent parts of the waves generated during times when the $m=5$ edge mode was particularly high. Assuming a group velocity of order 0.5 km/year, this would imply the $m=5$ pattern's amplitude varies with a period of order 30 years. This period is significantly longer than the full span of the Cassini mission and so the variations in the edge shape might have eluded detection thus far. Further study will therefore be needed to determine if the wave's detailed structure is consistent with the recent history of the gap edge.

\section{Candidates for additional axisymmetric density waves}
\label{others}

If axisymmetric density waves can be generated by combinations of normal modes on resonantly confined ring edges, then other edges confined by resonances could potentially produce axisymmetric waves. In Section~\ref{saturn} below we discuss additional candidate axisymmetric waves in Saturn's rings, while in Section~\ref{uranus} we briefly discuss possible analogs in the Uranian rings.

\subsection{Other axisymmetric waves in Saturn's rings}
\label{saturn}

A wavelet-based survey of the entire ring system failed to reveal any additional signals consistent with $m=0$ density waves that were as strong as the signal from the region interior to the Barnard Gap. This is not necessarily so surprising if such waves are indeed generated by mode mixing near edges, since most edges do not show the rich spectrum of modes seen on the inner edge of the Barnard Gap \citep{Nicholson14b, Nicholson14c, French16}. Instead, most inner edges of gaps are dominated by a single mode (usually $m=1$), which would probably not be ideal for generating the axisymmetric perturbations needed to produce a detectable wave. Even so, one might hope to see such waves generated at other resonantly-confined outer ring or inner gap edges, which are the closest analogs to the Barnard Gap. The other four edges in Saturn's rings that are clearly associated with satellite resonances are:
\begin{itemize}
\item The outer edge of the A ring, which is close to the 7:6 resonance with the coorbital moons Janus and Epimetheus \citep{SP09, EM16}
\item The inner edge of the Keeler Gap in the outer A ring, which is influenced by the 32:31 resonance with Prometheus \citep{Tajeddine17}.
\item The outer edge of the B ring, which is held in place by the 2:1 resonance with Mimas \citep{SP10, Nicholson14b}
\item The outer edge of the Dawes ringlet in the C ring, which is held in place by the 3:1 resonance with Mimas \citep{Nicholson14c}
\end{itemize}
Note that while other features are located close to strong satellite resonances --like the Laplace Ringlet in the Cassini Division, or the Bond and Colombo Ringlets in the C ring-- for these features the relevant resonances are located close to the center of the ringlet and so perturb the internal structure or global shape of the ringlet, rather than the position of one of its edges. We will not consider those features further here. 

Section \ref{aring} below briefly considers the edges of the A ring and the Keeler gap, neither of which seem to produce an axisymmetric density wave. However, as mentioned in the Introduction, both the Dawes ringlet and the outer B ring possess structures that could represent axisymmetric density waves. Section~\ref{bcedge} examines the occultation data in more detail, which provide evidence that these patterns are distorted by long-wavelength $m=1$ perturbations that complicate their interpretation. Section~\ref{bim} then discusses evidence from selected imaging sequences that the structures in the outer B ring do have some properties consistent with axisymmetric density waves.

\begin{figure}
\resizebox{3.4in}{!}{\includegraphics{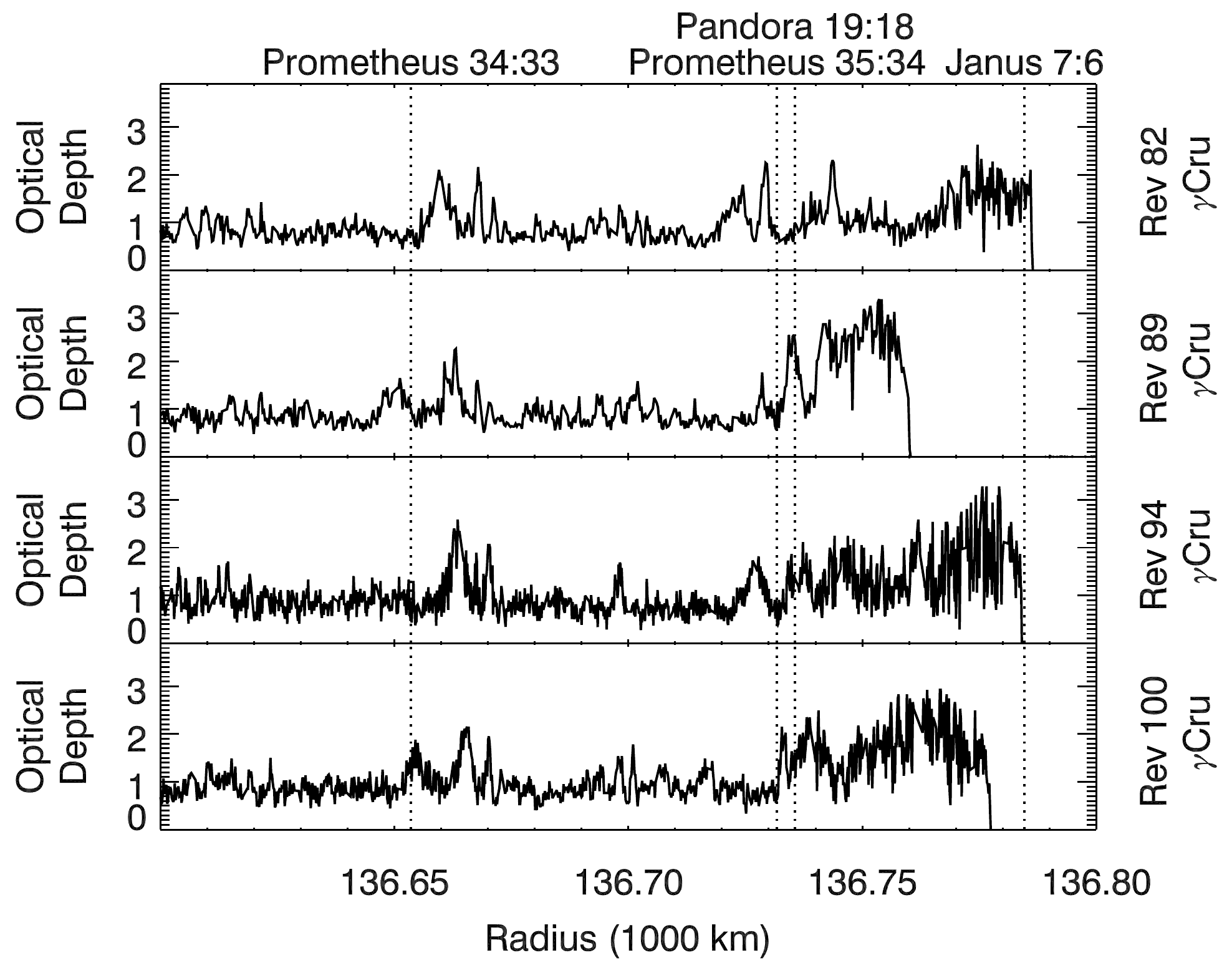}}
\caption{A few representative optical depth profiles of the A ring outer edge derived from stellar occultations obtained by the VIMS instrument onboard the Cassini spacecraft.}
\label{aedge}
\end{figure}

\begin{figure}
\resizebox{3.4in}{!}{\includegraphics{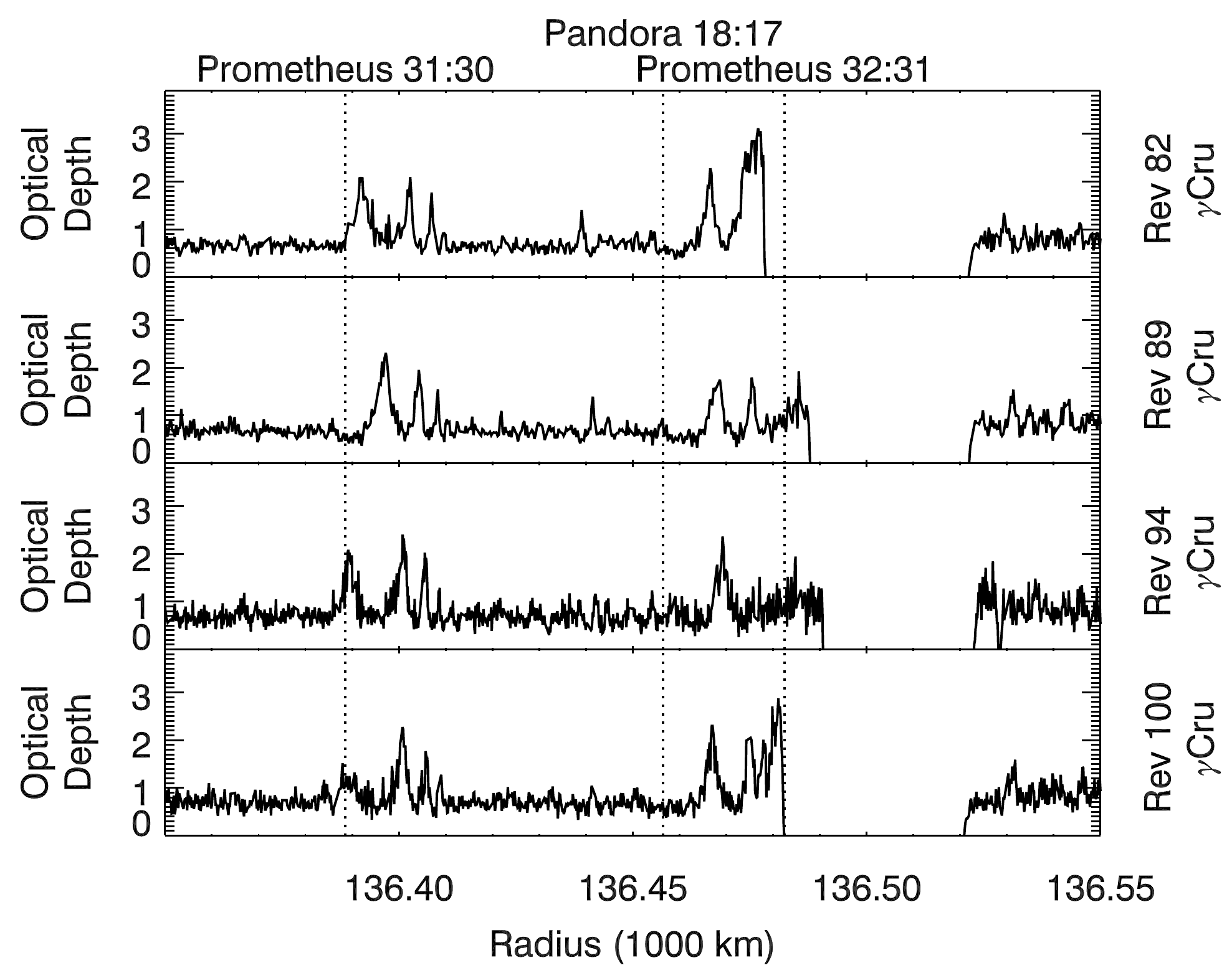}}
\caption{A few representative optical depth profiles of the Keeler Gap inner edge derived from stellar occultations obtained by the VIMS instrument onboard the Cassini spacecraft.}
\label{kedge}
\end{figure}

\subsubsection{No evidence for axisymmetric waves associated with edges in the A ring}
\label{aring}

Figures~\ref{aedge} and ~\ref{kedge} show representative profiles of the Keeler Gap and A-ring edges. While there are a small number of peaks near both of these edges, these can reasonably be attributed to outward-propagating density waves  generated by  the Pandora 19:18/Prometheus 35:34 or Pandora 18:17 resonances, respectively. Hence there is no evidence for axisymmetric density waves associated with either of these edges. 

Axisymmetric waves may not exist on these particular edges because their dynamical environments are somewhat different from that of the Barnard gap edge. While the Keeler gap is clearly perturbed by a resonance with Prometheus \citep{Tajeddine17}, this edge also falls close to the orbit of the much smaller moon Daphnis, and that moon may prevent the edge from developing a complex spectrum of normal modes. On the other hand, the outer edge of the A ring is perturbed by the co-orbital moons Janus and Epimetheus, whose periodic orbital changes clearly influence the shape of the edge over the course of a few years \citep{SP09, EM16}. A tesseral resonance with the planet may further complicate this situation \citep{EM16}. This edge's structure is therefore more time-dependent than the other edges considered here, which could inhibit the formation of additional edge modes and axisymmetric waves.

\begin{figure}
\resizebox{3.4in}{!}{\includegraphics{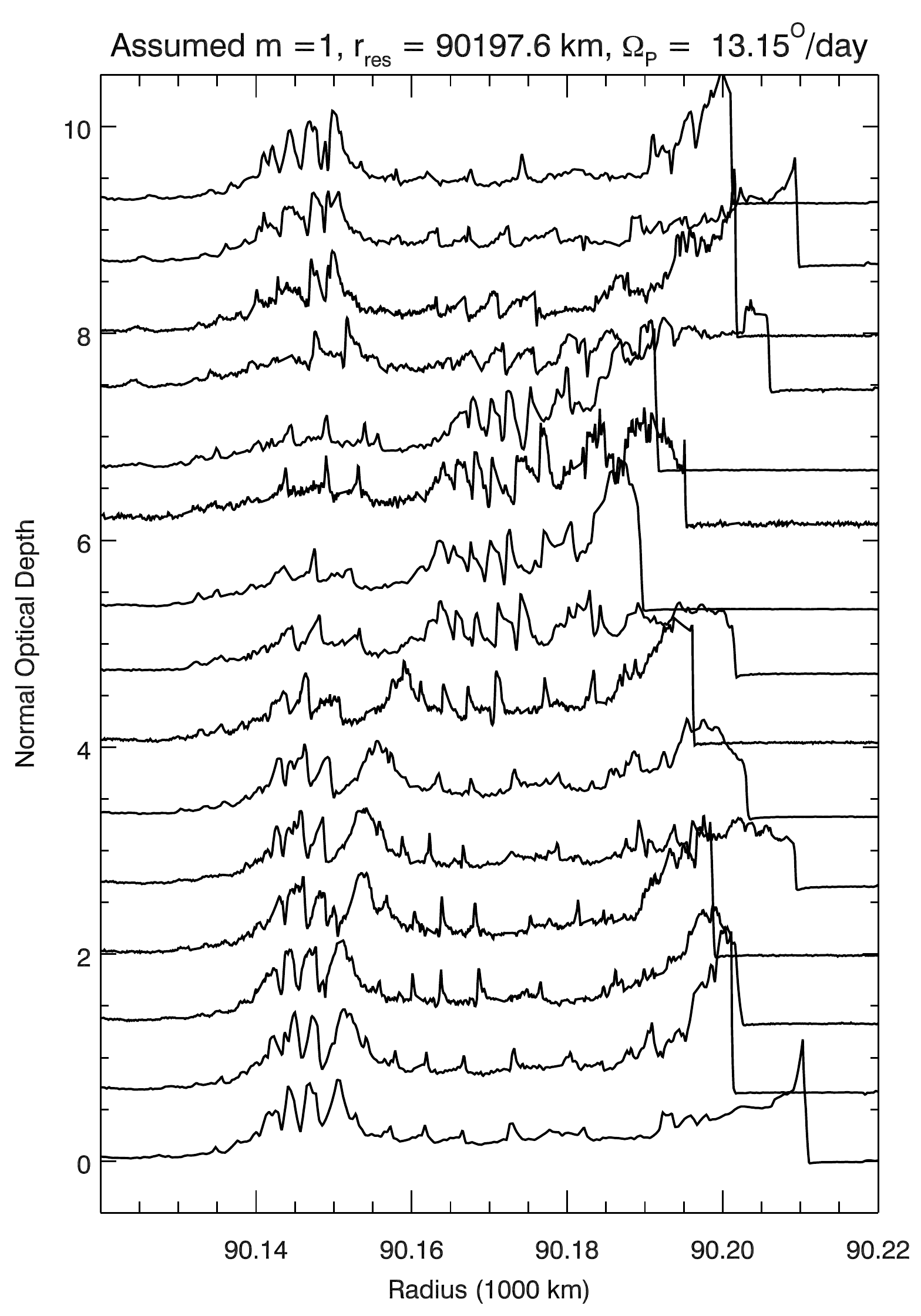}}
\caption{Plot showing a collection of profiles of the Dawes ringlet derived from occultations of the star $\gamma$ Crucis  obtained in 2008 and 2009. The profiles are sorted  by an $m=1$ phase computed using the indicated pattern speed. Note that both the shape and spacing of the periodic structure vary systematically from profile to profile.}
\label{dawes1}
\end{figure}

\subsubsection{Occultation profiles of the candidate axisymmetric waves in the Dawes ringlet and the outer B ring}
\label{bcedge}

As mentioned in the introduction to this paper, the Dawes ringlet does contain a series of peaks and dips that extend on either side of the Pandora 2:1 resonance, which is inconsistent with a wave launched from that resonance, and so could be due to a wave propagating inwards from the gap edge (see Figure~\ref{dawes}). Also, the complex structure of the B ring does include some periodic optical depth variations around 117,300 km that could represent parts of a similar wave (see Figure~\ref{bedge}). Furthermore, if axisymmetric waves are generated by interference among resonant perturbations and edge normal modes, then the Dawes ringlet and outer B ring are the most promising places to find additional examples of these waves. Both edges exhibit normal modes with amplitudes that are not much smaller than the $m=2$ patterns generated by the relevant satellite resonances \citep{Nicholson14b, Nicholson14c}. 

\begin{figure}
\resizebox{3.4in}{!}{\includegraphics{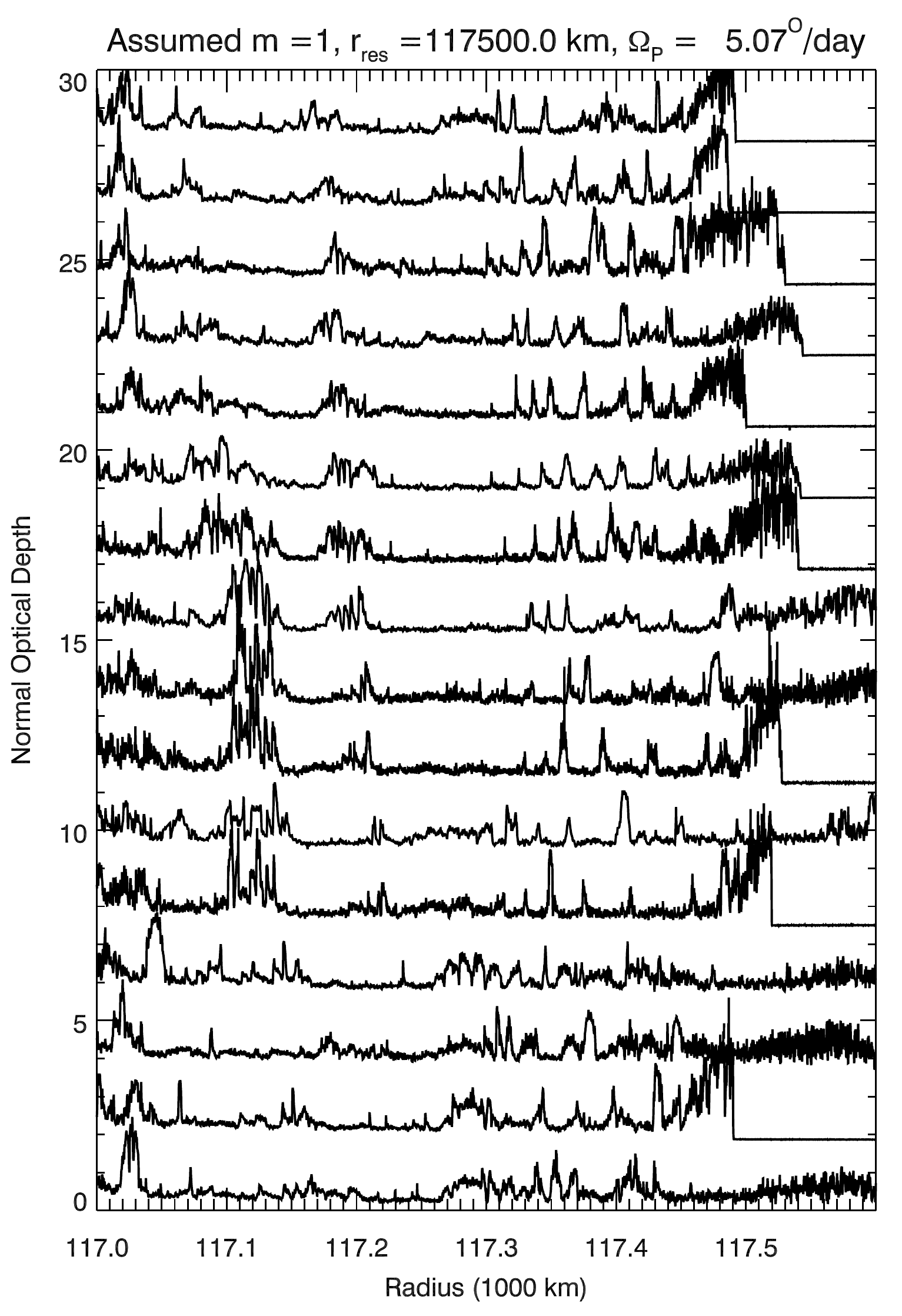}}
\caption{Plot showing a collection of profiles of the outer B ring oderived from occultations of the star $\gamma$ Crucis obtained in 2008 and 2009. The profiles are sorted in order of an $m=1$ phase computed using the indicated pattern speed. Note that the locations and extent of regions containing fine-scale structures vary systematically among these profiles.}
\label{bwave1}
\end{figure}

\begin{figure}
\resizebox{3.4in}{!}{\includegraphics{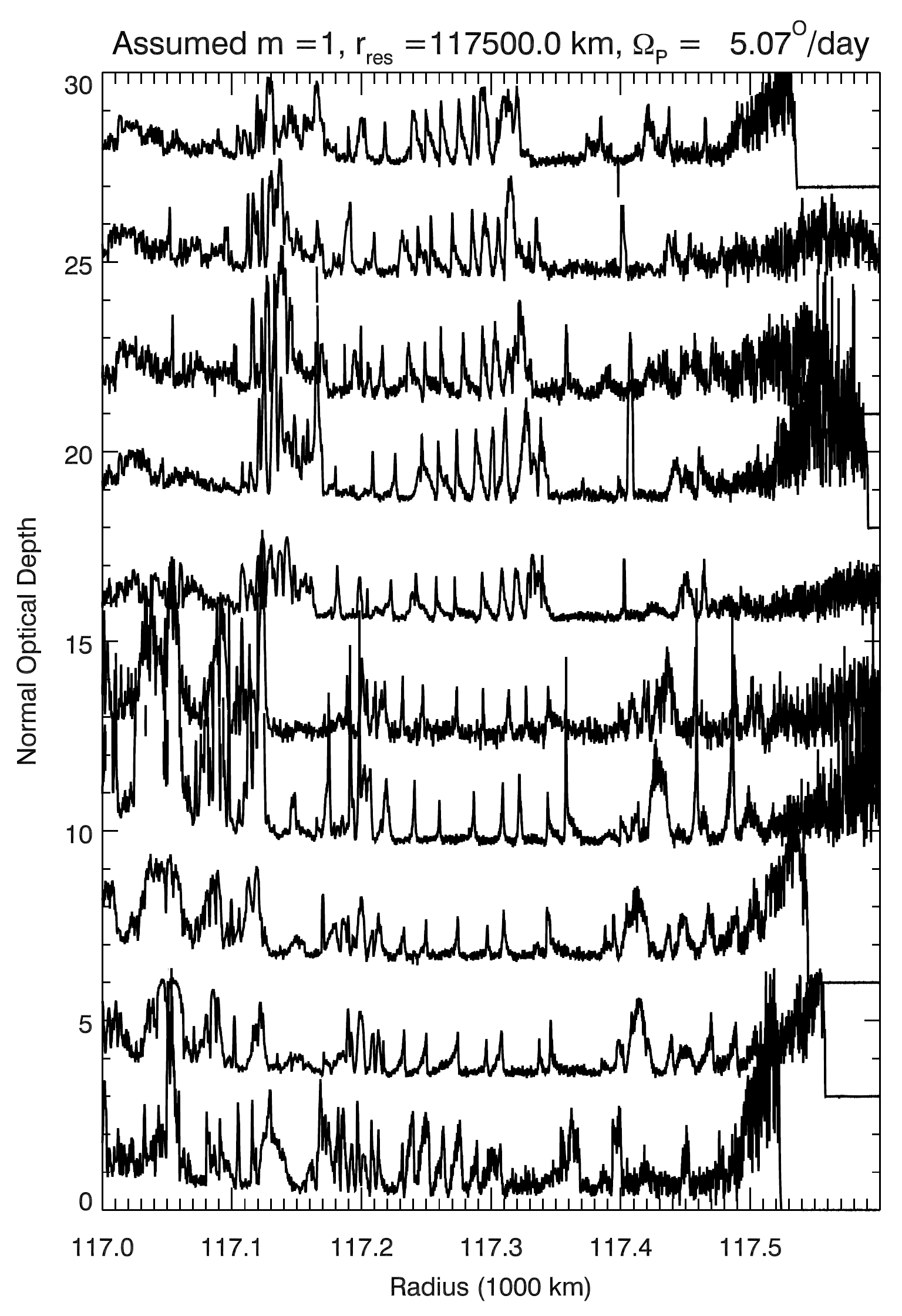}}
\caption{Plot showing a collection of profiles of the outer B ring derived from occultations of the star $\alpha$ Scorpii in 2016. The profiles are sorted in order of an $m=1$ phase computed using the indicated pattern speed. Note that the shapes and spacing of the peaks  vary systematically among these profiles.}
\label{bwave2}
\end{figure}

Applying the above wavelet-based analyses to the Dawes ringlet and outer B-ring failed to reveal any clear $m=0$ patterns in these regions. Closer inspection of the relevant patterns, however, reveals that the structures of both of these regions are probably much more heavily distorted by $m=1$ disturbances than is the region near the Barnard gap. Figures~\ref{dawes1}-\ref{bwave2} show multiple profiles of the Dawes ringlet and the outer B ring sorted by the phase of an $m=1$ pattern generated near the relevant edge (i.e. $\lambda-\dot{\varpi}(t-t_0)$). In Figure~\ref{dawes1} the profiles of the Dawes ringlet clearly indicate that both the shape and spacing of the periodic opacity peaks vary systematically from profile to profile. For the outer B ring, the situation is a bit more complicated. Figure~\ref{bwave1} shows profiles derived from occultations of the star $\gamma$ Crucis obtained in 2008-2009. Here there are systematic trends in the locations where small-scale periodic optical depth variations occur (e.g. the cluster of peaks around 117,200 km), but variations in wavelength are harder to discern. However, in later observations (shown in Figure~\ref{bwave2}) the periodic patterns are more prominent, and show variations in shape and spacing similar to those seen in the Dawes ringlet.

It is important to note that while the shape and spacing of the periodic optical depth variations are modulated by something with $m=1$, this does not mean that periodic structures are themselves $m=1$ spiral patterns. Recall that $m=1$ waves propagate outwards, so such waves would need to be generated interior to the relevant edges, and there are no known $m=1$ resonances that could excite waves in these regions. Instead, these periodic structures probably have another $m$-number and are distorted by the $m=1$ normal modes associated with the nearby edge.  Indeed, $m=1$ edge modes are expected to penetrate much further into the ring than other normal modes,\footnote{This is for the same reason that the radial wavelengths of $m=1$ density waves are much larger than those with other values of $m$.} so it is not unreasonable that the observed distortions are mostly $m=1$ \citep{SP10, Nicholson14b}. Also, the $m=1$ normal modes on the outer edges of the Dawes ringlet and the B-ring are over an order of magnitude larger than that found on the inner edge of the Barnard Gap, being $6.10\pm0.12$ km and $20.44\pm0.99$ km,. respectively \citep{Nicholson14b, Nicholson14c}, as opposed to $0.44\pm0.06$ km \citep{French16}. This would naturally explain why these distortions did not interfere with our wavelet analysis of the latter region.

In principle, the distortions in the pattern's wavelength induced by the $m=1$ edge modes could  be corrected for and removed using techniques such as normalizing the radius scale \citep{Graps95, French16c}. However, thus far we have not had any success in using these techniques to obtain a signal that can be clearly identified with wavelet tools like those discussed above. This is probably because the $m=1$ distortion is not simply a uniform telescoping of the entire region, but is instead a more complex distortion involving gradients in both the eccentricity and pericenter position. For example, if we examine the Dawes ringlet profiles in Figure~\ref{dawes1}, we can see that the wavelength of the periodic structures around 90,150 km reaches its maximum value when the pattern's wavelength around 90,170 km is near its minimum value. Thus we cannot use  the same sorts of wavelet analyses done above to ascertain if these patterns are truly due to a $m=0$ waves launched from the relevant edges.

\begin{figure}
\resizebox{3.4in}{!}{\includegraphics{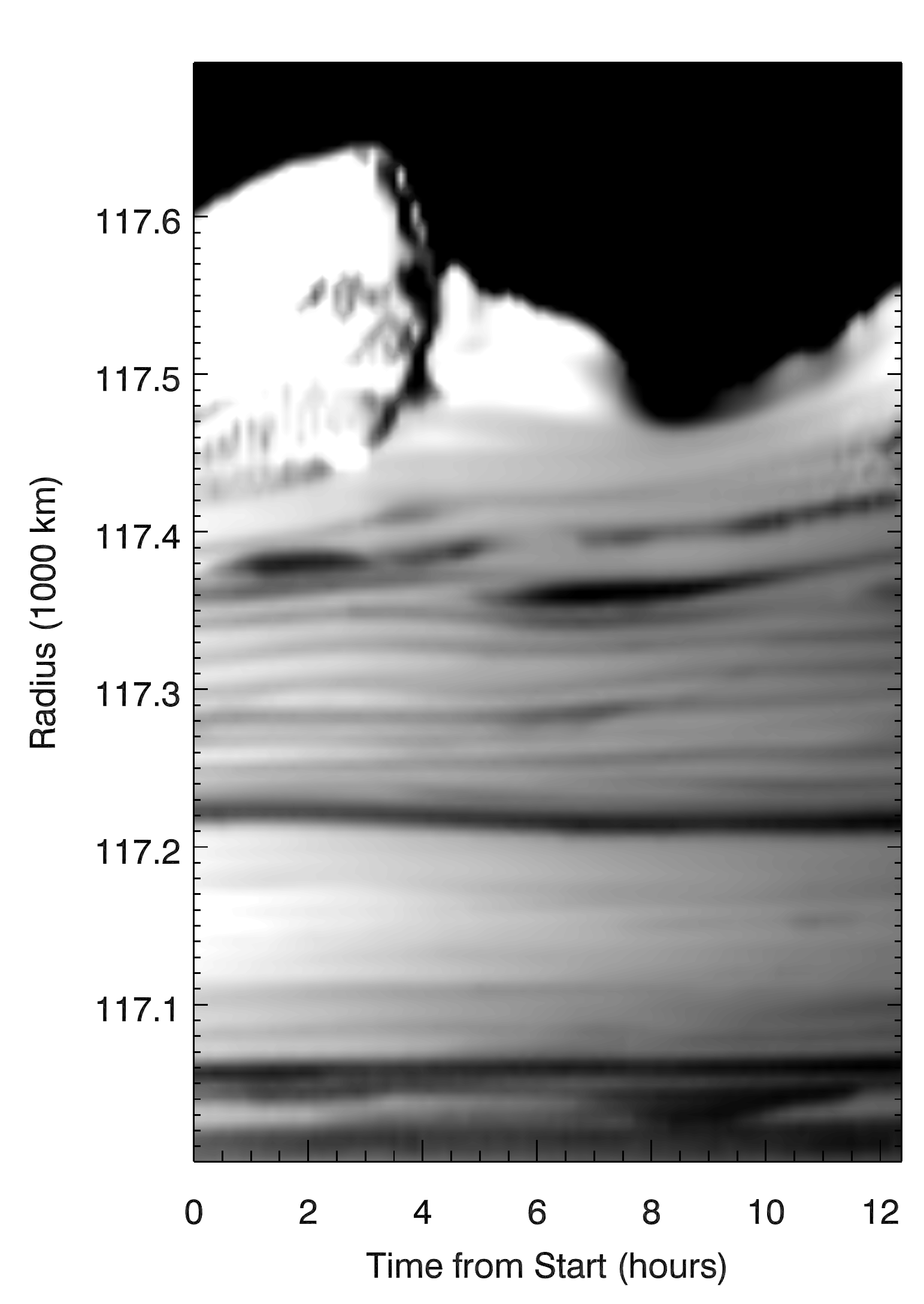}}
\caption{Image showing the brightness of the outer part of Saturn's B-ring as a function of radius and time at one inertial longitude, derived from a series of Cassini images obtained on day 198 of 2014 (i.e. near a maximum in the amplitude of the $m=2$ amplitude). Note the bright and dark bands between 117,200 km and 117,400 km. The phase speed of this pattern is consistent with an $m=0$ wave launched from the outer edge of the B ring.}
\label{bmovie}
\end{figure}

Rather than try to correct for these distortions, we could instead select observations made at nearly the same phase of the $m=1$ distortion. This strategy greatly reduces the number of possible comparisons among occultations, and unfortunately has not yet been productive for the occultation data.  However, circumstantial evidence for an axisymmetric density wave in the outer part of the B ring can be obtained from imaging data. 

\subsubsection{Evidence for an axisymmetric wave in the outer B-ring from images}
\label{bim}

For the occultations shown in Figure~\ref{bwave2}, the wavelength of the pattern  around 117,250 km is 10-30 km. which is large enough to be resolved in images obtained by the Narrow Angle Camera onboard the Cassini spacecraft \citep{Porco04}.  This instrument made several observations where it stared at a fixed longitude in the outer part of the B ring over the course of an orbital period (roughly 12 hours) and watched material rotate through the field of view. These particular observations are especially useful for this investigation because all the images were taken at the same inertial longitude.  Since the apsidal precession rates in this part of the ring are only about 5$^\circ$/day, this means they were all taken at basically the same phase of any $m=1$ structure. By contrast, the pattern speed for any other $m$ value is comparable to the orbital mean motion, and so changes associated with the changing phases of other patterns should be visible in these data. 

A thorough investigation of all the relevant imaging data is beyond the scope of this work, but we conducted a preliminary study of one observation sequence called BMOVIE from Rev 206, which consisted of 100 images (filenames N1784298322-N1784343322) of the unlit side of the outer B ring obtained on day 198 of 2014. We navigated these images with the appropriate SPICE kernels and adjusted estimates of the pointing so that the outer edge of the Jeffreys gap in the Cassini Division was at the expected position in all the images. We then averaged the brightness over longitude to generate a radial brightness profile from each image. Figure~\ref{bmovie} shows a mosaic of these brightness profiles that gives the brightness of the ring as a function of radius and time at the observed inertial longitude.  Note that the total timespan of the observation corresponds to approximately one local orbital period.

The most dramatic feature in this image is the variation in the location of the outer edge between 117,470 km and 117,630 km, which is dominated by perturbations driven by the Mimas 2:1 resonance. This is an $m=2$ pattern moving around the planet at around half the local orbital speed, so we only see a single maximum and minimum in the radial position over the course of the observation.

For the purposes of this analysis, however, the more interesting structures are the alternating bright and dark bands between 117,220 km and 117,400 km. These bands are in the same place as the periodic opacity variations seen in the occultation data in Figure~\ref{bwave2}, and so are almost certainly the same structure.  With the imaging data, we can clearly trace brightness maxima and minima over time, and while the bands seem to drift inwards and outwards at different times, the overall trend is for the bands to move outwards over time. This generally outwards motion of brightness maxima and minima is consistent with the expected behavior of crests and troughs of a density wave (recall from Section~\ref{background} that density waves always have a positive (outwards) phase velocity). The variations around this main trend, by contrast, are likely due to distortions in the ring associated with the various edge modes. Indeed, the wiggles in the positions of these dark bands seem to track the $m=2$ variations in the position of the edge.

If these periodic bands are in fact a density wave, we can use these data to constrain the value of $m$ for this pattern. Recall that the frequency $\omega$ of such a wave  can be written as a function of the pattern speed or of the resonant mean motion and epicyclic frequencies, so that the phase of the pattern $\phi_r+\phi_{\lambda t}$ can be written as (cf. Equations~\ref{phidef}-\ref{eqpat0}):

\begin{equation}
\phi \simeq \phi_r(r)+|m|\lambda-|m-1|n_L t
\end{equation}
This means that at a given radius and longitude, the phase should go through $|m-1|$ cycles in one orbital period, or equivalently, $|m-1|$ bright or dark bands should cross each radius in an orbit period. For the pattern in Figure~\ref{bmovie}, if we neglect the distortions associated with the edge modes, then we find that only a single bright or dark band crosses each radius in an orbit period. This means that $|m-1|$=1, which means that $m$ should be either 0 or 2. At first, $m=2$ seems like a more logical choice, given the strong $m=2$ Inner Lindblad Resonance near the edge, but an $m=2$ structure would propagate outwards, leading to smaller wavelengths closer to the edge, which is not the case.  Also, an $m=2$ resonant cavity mode should not extend this far inwards from the edge \citep{SP10, Nicholson14b}. Instead, the wavelength seems to decrease inwards, which is more consistent with an $m=0$ outer-Lindblad-resonance-like wave launched from the edge itself.\footnote{Recall that for density waves with $m<1$ the phase and group velocities are in opposite directions.} 

Again, further work will be needed to completely understand this structure in the outer B ring, but imaging data does provide evidence that an axisymmetric density wave is likely to be part of the structure in the outer B ring. Unfortunately, there are no movies with sufficient resolution of the Dawes ringlet to provide evidence that the periodic structures here are also an inward-propagating axisymmetric wave, but analogies with both the Barnard gap and the B ring make this likely. Also note that both the Dawes ringlet and B-ring edges show strong $m=1, 2$ and 3 normal modes in addition to the resonantly excited $m=2$ modes  \citep{Nicholson14b, Nicholson14c}, which could generate the $m=0$ wave through the same basic mechanism as described above for the Barnard Gap (in this case having the $m=2$ pattern generated by Mimas mix with normal modes with $m=1$ and 3). 

\bigskip

\subsection{Axisymmetric structures in the Uranian rings}
\label{uranus}

Additional axisymmetric structures might also be found in Uranus' narrow dense rings. Of course, the available data on these rings are much more limited, but there are at least four potential examples of resonantly-confined edges in Uranus' rings \citep{PG87, French91}:
\begin{itemize}
\item The outer edge of the $\epsilon$ ring is close to the  14:13 inner Lindblad resonance with Ophelia
\item The inner edge of the $\epsilon$ ring is close to the 24:25 outer Lindblad resonance with Cordelia
\item The outer edge of the $\delta$ ring is close to the 23:22 inner Lindblad  resonance with Cordelia
\item The inner edge of the $\gamma$ ring is close to the 6:5 inner Lindblad resonance with Ophelia
\end{itemize}
The $\eta$ ring is also perturbed by the 3:2 inner Lindblad resonance with Cressida \citep{Chancia17}, but this resonance is sufficiently distant from the ring that it affects the overall shape of the  ring, rather than a specific edge. 

Since axisymmetric density waves propagate inwards, the most likely places these waves would arise is from the outer edges of the $\epsilon$ and $\delta$ rings. Interestingly, inward-propagating density waves have been identified in both of these rings \citep{Horn88, YF92}. These waves were previously suggested to be generated by resonances with interior satellites, but the possibility that they are due to an axisymmetric wave generated by edge modes is worth exploring. One potential argument against such an interpretation is that the short wavelengths of these patterns suggest that each must represent a wave with a large $|m|$ \citep{Horn88, YF92}. 

Of course, the most famous axisymmetric structure in the Uranian rings is the $m=0$ ``breathing mode'' in the mean position of the $\gamma$ ring. Strangely, the nearest satellite resonance appears to be closer to the inner edge of this ring than its outer edge, so it may not be a perfect analog to the waves seen in Saturn's rings. However, there is evidence that the width of this ring has $m=4$ and $m=6$ patterns \citep{Showalter11}, and it could be that these represent suitable analogs of edge modes for this narrow ring. It may even be that this ring is so narrow that some modes primarily affect the width of the ring while others like the $m=0$ and $m=1$ primarily affect its mean radius \citep{Longaretti89}. Understanding how these modes could mix and interfere in this context could be a useful test for any model developed to explain the axisymmetric density waves in Saturn's rings.

\section{Summary}
\label{summary}

The basic results of the above investigation are:
\begin{itemize}
\item An inward-propagating axisymmetric density wave is being launched from the inner edge of the Barnard Gap in the Cassini Division.
\item This wave could be generated by interference between perturbations from the 5:4 resonance with Prometheus and normal modes on the gap edge.
\item Another example of an axisymmetric density wave probably exists in the outer B ring, being launched from that ring's outer edge near the Mimas 2:1 Lindblad Resonance.
\item A third example may be present in the Dawes ringlet in the C ring.
\item Additional examples may be present in the $\gamma$, $\delta$ and/or $\epsilon$ Uranian rings.
\end{itemize}
Future studies will be needed to develop a proper physical model of the mode mixing that could generate axisymmetric waves, to confirm or deny the existence of  the other possible waves associated with resonantly-confined edges, and to determine whether these waves play any significant role in angular momentum and energy transport near these edges \citep{Tajeddine17b} 

\section*{Acknowledgements}

This work was partially supported by a NASA Cassini Data Analysis Program Grant NNX14AD50G. We thank the Cassini project, as well as the VIMS and Imaging Teams for generating the data used in this analysis. We  also wish to acknowledge three undergraduate students, S. Graven, R. Miller, and R. Buckingham, whose examinations of the B ring helped put the axisymmetric structures in that region in context.

\section*{Appendix: A potential ambiguity in wave identifications}

While the 2008-2009 $\gamma$ Crucis occultation data shown in Figure~\ref{wavegam} are consistent with an $m=0$ density wave, for these particular occultations we cannot rule out the possibility that this is an $m=2$ bending wave.

Recall that the expected phase shifts for an $m=0$ density wave are $\phi_i(m=0) = -\kappa_L(t_i-t_0)$. The analogous expression for an $m=2$ bending wave is $\phi_{i,v}(m=2) = 2\lambda-(2n_L-\nu_L)(t_i-t_0)$, where $\nu_L$ is the vertical epicyclic frequency. Expressed in terms of the apsidal precession rate $\dot{\varpi}_L$ and the nodal regression rate $\dot{\Omega}_L$  these expressions become 

\begin{equation}
\phi_i(m=0) = -(n_L-\dot{\varpi}_L)(t_i-t_0)
\end{equation}and 

\begin{equation}
\phi_{i,v}(m=2) = 2\lambda-(n_L+\dot{\Omega}_L)(t_i-t_0).
\end{equation}
For all parts of Saturn's rings, $\dot{\Omega}_L \simeq -\dot{\varpi}_L$, and so the time-dependent parts of these two phases are nearly identical. This means that a set of occultations obtained at a single longitude $\lambda$ cannot distinguish between these two different types of waves because the predicted phase shifts differ by an additive constant $2\lambda$ that does not contribute to the wavelet power levels (In general, similar ambiguities can arise between spiral density waves with $|m|$ arms and bending waves with $|m+2|$ arms).

Fortunately, we can resolve this ambiguity and confirm our identification of this wave by considering the full set of occultations, which cut this region at different longitudes as well as different times (they also had a range of opening angles, which also influence the inferred phase of any potential bending wave). This extended data set yields no signal for the $m=2$ vertical wave option but is consistent with an $m=0$ density wave (see Figure~\ref{wavefull}).


\begin{thebibliography}{}
\makeatletter
\relax
\def\mn@urlcharsother{\let\do\@makeother \do\$\do\&\do\#\do\^\do\_\do\%\do\~}
\def\mn@doi{\begingroup\mn@urlcharsother \@ifnextchar [ {\mn@doi@}
  {\mn@doi@[]}}
\def\mn@doi@[#1]#2{\def\@tempa{#1}\ifx\@tempa\@empty \href
  {http://dx.doi.org/#2} {doi:#2}\else \href {http://dx.doi.org/#2} {#1}\fi
  \endgroup}
\def\mn@eprint#1#2{\mn@eprint@#1:#2::\@nil}
\def\mn@eprint@arXiv#1{\href {http://arxiv.org/abs/#1} {{\tt arXiv:#1}}}
\def\mn@eprint@dblp#1{\href {http://dblp.uni-trier.de/rec/bibtex/#1.xml}
  {dblp:#1}}
\def\mn@eprint@#1:#2:#3:#4\@nil{\def\@tempa {#1}\def\@tempb {#2}\def\@tempc
  {#3}\ifx \@tempc \@empty \let \@tempc \@tempb \let \@tempb \@tempa \fi \ifx
  \@tempb \@empty \def\@tempb {arXiv}\fi \@ifundefined
  {mn@eprint@\@tempb}{\@tempb:\@tempc}{\expandafter \expandafter \csname
  mn@eprint@\@tempb\endcsname \expandafter{\@tempc}}}

\bibitem[\protect\citeauthoryear{{Acton}}{{Acton}}{1996}]{Acton96}
{Acton} C.~H.,  1996, \mn@doi [\planss] {10.1016/0032-0633(95)00107-7}, \href
  {http://adsabs.harvard.edu/abs/1996P%26SS...44...65A} {44, 65}

\bibitem[\protect\citeauthoryear{{Brown} et~al.,}{{Brown}
  et~al.}{2004}]{Brown04}
{Brown} R.~H.,  et~al., 2004, \mn@doi [Space Science Reviews]
  {10.1007/s11214-004-1453-x}, \href
  {http://adsabs.harvard.edu/abs/2004SSRv..115..111B} {115, 111}

\bibitem[\protect\citeauthoryear{{Chancia}, {Hedman}  \& {French}}{{Chancia}
  et~al.}{2017}]{Chancia17}
{Chancia} R.~O.,  {Hedman} M.~M.,   {French} R.~G.,  2017, \mn@doi [AJ]
  {10.3847/1538-3881/aa880e}, \href
  {http://adsabs.harvard.edu/abs/2017AJ....154..153C} {154, 153}

\bibitem[\protect\citeauthoryear{{Colwell}, {Nicholson}, {Tiscareno}, {Murray},
  {French}  \& {Marouf}}{{Colwell} et~al.}{2009}]{Colwell09}
{Colwell} J.~E.,  {Nicholson} P.~D.,  {Tiscareno} M.~S.,  {Murray} C.~D.,
  {French} R.~G.,   {Marouf} E.~A.,  2009, in {Dougherty, M.~K., Esposito,
  L.~W., \& Krimigis, S.~M.} ed., Saturn from Cassini-Huygens. Cmabridge U., pp
  375--412, \mn@doi{10.1007/978-1-4020-9217-6_13}

\bibitem[\protect\citeauthoryear{{Cuzzi}, {Lissauer}  \& {Shu}}{{Cuzzi}
  et~al.}{1981}]{Cuzzi81}
{Cuzzi} J.~N.,  {Lissauer} J.~J.,   {Shu} F.~H.,  1981, \mn@doi [Nature]
  {10.1038/292703a0}, \href {http://adsabs.harvard.edu/abs/1981Natur.292..703C}
  {292, 703}

\bibitem[\protect\citeauthoryear{{El Moutamid} et~al.,}{{El Moutamid}
  et~al.}{2016}]{EM16}
{El Moutamid} M.,  et~al., 2016, \mn@doi [Icarus]
  {10.1016/j.icarus.2015.10.025}, \href
  {http://adsabs.harvard.edu/abs/2016Icar..279..125E} {279, 125}

\bibitem[\protect\citeauthoryear{{Esposito}}{{Esposito}}{2010}]{Esposito10}
{Esposito} L.~W.,  2010, \mn@doi [Annual Review of Earth and Planetary
  Sciences] {10.1146/annurev-earth-040809-152339}, \href
  {http://adsabs.harvard.edu/abs/2010AREPS..38..383E} {38, 383}

\bibitem[\protect\citeauthoryear{{Esposito}, {Ocallaghan}  \&
  {West}}{{Esposito} et~al.}{1983}]{Esposito83}
{Esposito} L.~W.,  {Ocallaghan} M.,   {West} R.~A.,  1983, \mn@doi [Icarus]
  {10.1016/0019-1035(83)90165-3}, \href
  {http://adsabs.harvard.edu/abs/1983Icar...56..439E} {56, 439}

\bibitem[\protect\citeauthoryear{{French}, {Nicholson}, {Porco}  \&
  {Marouf}}{{French} et~al.}{1991}]{French91}
{French} R.~G.,  {Nicholson} P.~D.,  {Porco} C.~C.,   {Marouf} E.~A.,  1991,
  {Dynamics and structure of the Uranian rings}.
pp 327--409

\bibitem[\protect\citeauthoryear{{French} et~al.,}{{French}
  et~al.}{1993}]{French93}
{French} R.~G.,  et~al., 1993, \mn@doi [Icarus] {10.1006/icar.1993.1066}, \href
  {http://adsabs.harvard.edu/abs/1993Icar..103..163F} {103, 163}

\bibitem[\protect\citeauthoryear{{French}, {Marouf}, {Rappaport}  \&
  {McGhee}}{{French} et~al.}{2010}]{French10}
{French} R.~G.,  {Marouf} E.~A.,  {Rappaport} N.~J.,   {McGhee} C.~A.,  2010,
  \mn@doi [AJ] {10.1088/0004-6256/139/4/1649}, \href
  {http://adsabs.harvard.edu/abs/2010AJ....139.1649F} {139, 1649}

\bibitem[\protect\citeauthoryear{{French} et~al.,}{{French}
  et~al.}{2017}]{French17}
{French} R.~G.,  et~al., 2017, \mn@doi [Icarus] {10.1016/j.icarus.2017.02.007},
  \href {http://adsabs.harvard.edu/abs/2017Icar..290...14F} {290, 14}

\bibitem[\protect\citeauthoryear{{French}, {Nicholson}, {McGhee-French},
  {Lonergan}, {Sepersky}, {Hedman}, {Marouf}  \& {Colwell}}{{French}
  et~al.}{2016a}]{French16}
{French} R.~G.,  {Nicholson} P.~D.,  {McGhee-French} C.~A.,  {Lonergan} K.,
  {Sepersky} T.,  {Hedman} M.~M.,  {Marouf} E.~A.,   {Colwell} J.~E.,  2016a,
  \mn@doi [Icarus] {10.1016/j.icarus.2016.03.017}, \href
  {http://adsabs.harvard.edu/abs/2016Icar..274..131F} {274, 131}

\bibitem[\protect\citeauthoryear{{French}, {Nicholson}, {Hedman}, {Hahn},
  {McGhee-French}, {Colwell}, {Marouf}  \& {Rappaport}}{{French}
  et~al.}{2016b}]{French16c}
{French} R.~G.,  {Nicholson} P.~D.,  {Hedman} M.~M.,  {Hahn} J.~M.,
  {McGhee-French} C.~A.,  {Colwell} J.~E.,  {Marouf} E.~A.,   {Rappaport}
  N.~J.,  2016b, \mn@doi [\icarus] {10.1016/j.icarus.2015.08.020}, \href
  {http://adsabs.harvard.edu/abs/2016Icar..279...62F} {279, 62}

\bibitem[\protect\citeauthoryear{{Goldreich} \& {Tremaine}}{{Goldreich} \&
  {Tremaine}}{1982}]{GT82}
{Goldreich} P.,  {Tremaine} S.,  1982, \mn@doi [\araa]
  {10.1146/annurev.aa.20.090182.001341}, \href
  {http://adsabs.harvard.edu/abs/1982ARA%26A..20..249G} {20, 249}

\bibitem[\protect\citeauthoryear{{Graps}, {Showalter}, {Lissauer}  \&
  {Kary}}{{Graps} et~al.}{1995}]{Graps95}
{Graps} A.~L.,  {Showalter} M.~R.,  {Lissauer} J.~J.,   {Kary} D.~M.,  1995,
  \mn@doi [AJ] {10.1086/117451}, \href
  {http://adsabs.harvard.edu/abs/1995AJ....109.2262G} {109, 2262}

\bibitem[\protect\citeauthoryear{{Hedman} \& {Nicholson}}{{Hedman} \&
  {Nicholson}}{2014}]{HN14}
{Hedman} M.~M.,  {Nicholson} P.~D.,  2014, \mn@doi [MNRAS]
  {10.1093/mnras/stu1503}, \href
  {http://adsabs.harvard.edu/abs/2014MNRAS.444.1369H} {444, 1369}

\bibitem[\protect\citeauthoryear{{Hedman} \& {Nicholson}}{{Hedman} \&
  {Nicholson}}{2016}]{HN16}
{Hedman} M.~M.,  {Nicholson} P.~D.,  2016, \mn@doi [Icarus]
  {10.1016/j.icarus.2016.01.007}, \href
  {http://adsabs.harvard.edu/abs/2016Icar..279..109H} {279, 109}

\bibitem[\protect\citeauthoryear{{Hedman} et~al.,}{{Hedman}
  et~al.}{2010}]{Hedman10}
{Hedman} M.~M.,  et~al., 2010, \mn@doi [\aj] {10.1088/0004-6256/139/1/228},
  \href {http://adsabs.harvard.edu/abs/2010AJ....139..228H} {139, 228}

\bibitem[\protect\citeauthoryear{{Horn}, {Lane}, {Yanamandra-Fisher}  \&
  {Esposito}}{{Horn} et~al.}{1988}]{Horn88}
{Horn} L.~J.,  {Lane} A.~L.,  {Yanamandra-Fisher} P.~A.,   {Esposito} L.~W.,
  1988, \mn@doi [Icarus] {10.1016/0019-1035(88)90016-4}, \href
  {http://adsabs.harvard.edu/abs/1988Icar...76..485H} {76, 485}

\bibitem[\protect\citeauthoryear{{Jacobson}, {Spitale}, {Porco}, {Beurle},
  {Cooper}, {Evans}  \& {Murray}}{{Jacobson} et~al.}{2008}]{Jacobson08}
{Jacobson} R.~A.,  {Spitale} J.,  {Porco} C.~C.,  {Beurle} K.,  {Cooper} N.~J.,
   {Evans} M.~W.,   {Murray} C.~D.,  2008, \mn@doi [\aj]
  {10.1088/0004-6256/135/1/261}, \href
  {http://adsabs.harvard.edu/abs/2008AJ....135..261J} {135, 261}

\bibitem[\protect\citeauthoryear{{Longaretti}}{{Longaretti}}{1989}]{Longaretti89}
{Longaretti} P.-Y.,  1989, \mn@doi [Icarus] {10.1016/0019-1035(89)90039-0},
  \href {http://adsabs.harvard.edu/abs/1989Icar...82..281L} {82, 281}
  
  \bibitem[\protect\citeauthoryear{{Nicholson}, {Cooke}  \& {Pelton}}{{Nicholson}
  et~al.}{1990}]{NCP90}
{Nicholson} P.~D.,  {Cooke} M.~L.,   {Pelton} E.,  1990, \mn@doi [\aj]
  {10.1086/115601}, \href {http://adsabs.harvard.edu/abs/1990AJ....100.1339N}
  {100, 1339}

\bibitem[\protect\citeauthoryear{{Nicholson}, {French}, {Hedman}, {Marouf}  \&
  {Colwell}}{{Nicholson} et~al.}{2014a}]{Nicholson14b}
{Nicholson} P.~D.,  {French} R.~G.,  {Hedman} M.~M.,  {Marouf} E.~A.,
  {Colwell} J.~E.,  2014a, \mn@doi [Icarus] {10.1016/j.icarus.2013.09.002},
  \href {http://adsabs.harvard.edu/abs/2014Icar..227..152N} {227, 152}

\bibitem[\protect\citeauthoryear{{Nicholson}, {French}, {McGhee-French},
  {Hedman}, {Marouf}, {Colwell}, {Lonergan}  \& {Sepersky}}{{Nicholson}
  et~al.}{2014b}]{Nicholson14c}
{Nicholson} P.~D.,  {French} R.~G.,  {McGhee-French} C.~A.,  {Hedman} M.~M.,
  {Marouf} E.~A.,  {Colwell} J.~E.,  {Lonergan} K.,   {Sepersky} T.,  2014b,
  \mn@doi [Icarus] {10.1016/j.icarus.2014.06.024}, \href
  {http://adsabs.harvard.edu/abs/2014Icar..241..373N} {241, 373}

\bibitem[\protect\citeauthoryear{{Porco} \& {Goldreich}}{{Porco} \&
  {Goldreich}}{1987}]{PG87}
{Porco} C.~C.,  {Goldreich} P.,  1987, \mn@doi [\aj] {10.1086/114354}, \href
  {http://adsabs.harvard.edu/abs/1987AJ.....93..724P} {93, 724}

\bibitem[\protect\citeauthoryear{{Porco} et~al.,}{{Porco}
  et~al.}{2004}]{Porco04}
{Porco} C.~C.,  et~al., 2004, \mn@doi [SSR] {10.1007/s11214-004-1456-7}, \href
  {http://adsabs.harvard.edu/abs/2004SSRv..115..363P} {115, 363}

\bibitem[\protect\citeauthoryear{{Porco} et~al.,}{{Porco}
  et~al.}{2005}]{Porco05}
{Porco} C.~C.,  et~al., 2005, \mn@doi [Science] {10.1126/science.1108056},
  \href {http://adsabs.harvard.edu/abs/2005Sci...307.1226P} {307, 1226}

\bibitem[\protect\citeauthoryear{{Porco}, {Thomas}, {Weiss}  \&
  {Richardson}}{{Porco} et~al.}{2007}]{Porco07}
{Porco} C.~C.,  {Thomas} P.~C.,  {Weiss} J.~W.,   {Richardson} D.~C.,  2007,
  \mn@doi [Science] {10.1126/science.1143977}, \href
  {http://adsabs.harvard.edu/abs/2007Sci...318.1602P} {318, 1602}

\bibitem[\protect\citeauthoryear{{Showalter}}{{Showalter}}{2011}]{Showalter11}
{Showalter} M.~R.,  2011, in EPSC-DPS Joint Meeting 2011. p.~1224

\bibitem[\protect\citeauthoryear{{Shu}}{{Shu}}{1984}]{Shu84}
{Shu} F.~H.,  1984, in {Greenberg} R.,  {Brahic} A.,  eds, Planetary Rings. U.
  of Arizona Press, pp 513--561

\bibitem[\protect\citeauthoryear{{Spitale} \& {Porco}}{{Spitale} \&
  {Porco}}{2009}]{SP09}
{Spitale} J.~N.,  {Porco} C.~C.,  2009, \mn@doi [AJ]
  {10.1088/0004-6256/138/5/1520}, \href
  {http://adsabs.harvard.edu/abs/2009AJ....138.1520S} {138, 1520}

\bibitem[\protect\citeauthoryear{{Spitale} \& {Porco}}{{Spitale} \&
  {Porco}}{2010}]{SP10}
{Spitale} J.~N.,  {Porco} C.~C.,  2010, \mn@doi [AJ]
  {10.1088/0004-6256/140/6/1747}, \href
  {http://adsabs.harvard.edu/abs/2010AJ....140.1747S} {140, 1747}

\bibitem[\protect\citeauthoryear{{Tajeddine}, {Nicholson}, {Longaretti}, {El
  Moutamid}  \& {Burns}}{{Tajeddine} et~al.}{2017a}]{Tajeddine17b}
{Tajeddine} R.,  {Nicholson} P.~D.,  {Longaretti} P.-Y.,  {El Moutamid} M.,
  {Burns} J.~A.,  2017a, \mn@doi [ApJS] {10.3847/1538-4365/aa8c09}, \href
  {http://adsabs.harvard.edu/abs/2017ApJS..232...28T} {232, 28}

\bibitem[\protect\citeauthoryear{{Tajeddine}, {Nicholson}, {Tiscareno},
  {Hedman}, {Burns}  \& {Moutamid}}{{Tajeddine} et~al.}{2017b}]{Tajeddine17}
{Tajeddine} R.,  {Nicholson} P.~D.,  {Tiscareno} M.~S.,  {Hedman} M.~M.,
  {Burns} J.~A.,   {Moutamid} M.~E.,  2017b, \mn@doi [Icarus]
  {10.1016/j.icarus.2017.02.002}, \href
  {http://adsabs.harvard.edu/abs/2017Icar..289...80T} {289, 80}

\bibitem[\protect\citeauthoryear{{Tiscareno}, {Nicholson}, {Burns}, {Hedman}
  \& {Porco}}{{Tiscareno} et~al.}{2006}]{Tiscareno06}
{Tiscareno} M.~S.,  {Nicholson} P.~D.,  {Burns} J.~A.,  {Hedman} M.~M.,
  {Porco} C.~C.,  2006, \mn@doi [ApJL] {10.1086/509120}, \href
  {http://adsabs.harvard.edu/abs/2006ApJ...651L..65T} {651, L65}

\bibitem[\protect\citeauthoryear{{Tiscareno}, {Burns}, {Nicholson}, {Hedman}
  \& {Porco}}{{Tiscareno} et~al.}{2007}]{Tiscareno07}
{Tiscareno} M.~S.,  {Burns} J.~A.,  {Nicholson} P.~D.,  {Hedman} M.~M.,
  {Porco} C.~C.,  2007, \mn@doi [Icarus] {10.1016/j.icarus.2006.12.025}, \href
  {http://adsabs.harvard.edu/abs/2007Icar..189...14T} {189, 14}

\bibitem[\protect\citeauthoryear{{Tiscareno}, {Hedman}, {Burns}, {Weiss}  \&
  {Porco}}{{Tiscareno} et~al.}{2013}]{Tiscareno13}
{Tiscareno} M.~S.,  {Hedman} M.~M.,  {Burns} J.~A.,  {Weiss} J.~W.,   {Porco}
  C.~C.,  2013, \mn@doi [Icarus] {10.1016/j.icarus.2013.02.026}, \href
  {http://adsabs.harvard.edu/abs/2013Icar..224..201T} {224, 201}

\bibitem[\protect\citeauthoryear{{Torrence} \& {Compo}}{{Torrence} \&
  {Compo}}{1998}]{TC98}
{Torrence} C.,  {Compo} G.~P.,  1998, \mn@doi [Bulletin of the American
  Meteorological Society] {10.1175/1520-0477(1998)079<0061:APGTWA>2.0.CO;2},
  \href {http://adsabs.harvard.edu/abs/1998BAMS...79...61T} {79, 61}

\bibitem[\protect\citeauthoryear{{Yanamandra-Fisher}}{{Yanamandra-Fisher}}{1992}]{YF92}
{Yanamandra-Fisher} P.~A.,  1992, \mn@doi [Advances in Space Research]
  {10.1016/0273-1177(92)90431-V}, \href
  {http://adsabs.harvard.edu/abs/1992AdSpR..12..149Y} {12, 149}

\bibitem[\protect\citeauthoryear{{Yoder}, {Colombo}, {Synnott}  \&
  {Yoder}}{{Yoder} et~al.}{1983}]{Yoder83}
{Yoder} C.~F.,  {Colombo} G.,  {Synnott} S.~P.,   {Yoder} K.~A.,  1983, \mn@doi
  [Icarus] {10.1016/0019-1035(83)90207-5}, \href
  {http://adsabs.harvard.edu/abs/1983Icar...53..431Y} {53, 431}

\makeatother
\end{thebibliography}

\end{document}